\def\be{\begin{equation}}
\def\ee{\end{equation}}
\def\bea{\begin{eqnarray*}}
\def\eea{\end{eqnarray*}}
\newcommand{\degree}{\ensuremath{^\circ}}

\def\url#1{\expandafter\string\csname #1\endcasname}
\def\H2{$H_2$}

\def\kms{km s$^{-1}$}
\def\Spitzer{\textit{Spitzer}}

\def\Herschel{\textit{Herschel}}
\def\Planck{\textit{Planck}}

\def\SPT{\textit{SPT}}

\def\WISE{\textit{WISE}}

\documentclass[useAMS,usenatbib]{mn2e}
\usepackage{subfigure}
\usepackage{natbib}
\usepackage{graphicx}
\usepackage{xcolor}
\usepackage{flafter}
\usepackage{lscape}
\usepackage{amssymb}
\usepackage{times}
\begin{document}
\title[Luminous High Redshift IR Galaxies Detected by \textit{Planck}]{Early Science with the Large Millimeter Telescope: Observations of Extremely Luminous High-z Sources Identified by \textit{Planck}}

\author[K. C.~Harrington et al.]{K.~C. ~Harrington$^1$,
Min S.~Yun$^1$,
R.~Cybulski$^1$,
G. W.~Wilson$^1$,
I.~Aretxaga$^2$,
\newauthor
M. Chavez$^2$, 
V.~De~la~Luz$^3$, 
N.~Erickson$^1$,
D.~Ferrusca$^2$, 
A.~D.~Gallup$^1$,
D. H.~Hughes$^2$, 
\newauthor
A.~Monta{\~n}a$^{2,4}$, 
G.~Narayanan$^1$,
D.~S\'{a}nchez-Arg\"{u}elles$^2$,
F.~P.~Schloerb$^1$,
\newauthor
K.~Souccar$^1$,
E. Terlevich$^2$,
R. Terlevich$^{2,5}$, 
M.~Zeballos$^2$, 
J. ~A.~Zavala$^2$ 
\\
$^1$Department of Astronomy, University of Massachusetts, Amherst, MA 01003, USA\\
$^2$Instituto Nacional de Astrof\'{i}sica, \'{O}ptica y Electr\'{o}nica, Tonantzintla, Puebla, M\'{e}xico\\
$^3$CONACYT Research Fellow - SCiESMEX, Instituto de Geof\'{i}sica, Unidad Michoac\'{a}n, Universidad Nacional Aut\'{o}noma de M\'{e}xico, Morelia,
Michoac\'{a}n, M\'{e}xico CP 58190\\
$^4$Consejo Nacional de Ciencia y Tecnolog\'{i}a, Av. Insurgentes Sur 1582, Col. Cr\'{e}dito Constructor, Del. Benito Ju\'{a}rez, C.P.: 03940, M\'{e}xico, D.F.\\
$^5$Institute of Astronomy, University of Cambridge, Madingley Road, Cambridge CB3 0HA, UK\\
}

\date{\today}
\pagerange{\pageref{firstpage}--\pageref{lastpage}} \pubyear{2016}
\maketitle
\label{firstpage}

\begin{abstract}
We present 8.5\arcsec\ resolution 1.1mm continuum imaging and CO spectroscopic redshift measurements of eight extremely bright submillimetre galaxies identified from the \Planck\ and \Herschel\ surveys, taken with the Large Millimeter Telescope's AzTEC and Redshift Search Receiver instruments. We compiled a candidate list of high redshift galaxies by cross-correlating the \textit{Planck Surveyor} mission's highest frequency channel (857 GHz, FWHM = 4.5\arcmin) with the archival \Herschel\ Spectral and Photometric Imaging Receiver (SPIRE) imaging data, and requiring the presence of a unique, single \Herschel\ counterpart within the 150\arcsec\ search radius of the \Planck\ source positions with 350 \micron\ flux density larger than 100 mJy, excluding known blazars and foreground galaxies. All eight candidate objects observed are detected in 1.1mm continuum by AzTEC bolometer camera, and at least one CO line is detected in all cases with a spectroscopic redshift between $1.3 < z_{CO} < 3.3$.   Their infrared spectral energy distributions mapped using the \Herschel\ and AzTEC photometry are consistent with cold dust emission with characteristic temperature between $T_d = 43$ K and 84 K.  With apparent infrared luminosity of  up to $L_{IR} = 3\times10^{14} \mu^{-1} L_\odot$, they are some of the most luminous galaxies ever found (with yet unknown gravitational magnification factor $\mu$).  The analysis of their spectral energy distributions (SEDs) suggests that star formation is powering the bulk of their extremely large IR luminosities. Derived molecular gas masses of $ M_{H2}=(0.6-7.8)\times 10^{11}  M_\odot$ (for $\mu\approx10$) also make them some of the most gas-rich high redshift galaxies ever detected. 
\end{abstract}

\section{Introduction}

	Submillimetre bright galaxies (``SMGs" hereafter) have emerged as an important component in our understanding of the evolution of galaxy formation in dusty, high-mass, starburst systems \citep[see the reviews by][]{blain02,casey14}. Studies of SMGs capture the dust enshrouded young stars and the supernovae remnants of massive early type OB stars within the cores of Giant Molecular Clouds (GMCs), revealing the nature of stellar assembly in the early universe through sub-mm through radio observations of thermal emission and molecular emission lines. SMG studies contribute to comprehensive multi-wavelength analyses of growth and evolution of some of the earliest forming galaxies emitting in the rest-frame IR \citep[see ][and references therein]{blain02}. 	The epoch of massive galaxy formation and growth encompasses a rich period ($1< z< 3.5$) of star formation, whereby parsec--to--kiloparsec scale GMCs are changing their distribution in the galaxy throughout various stages of merging, gas accretion and star formation. This epoch consists of active star formation rates that are particularly bright in the infrared to sub-millimeter ranges.  	
    	The \Planck\  mission's all-sky survey \citep{planck1} has proven to be useful in detecting highly luminous galaxies ($L_{IR} > 10^{12-14} L_\odot$) over a wide range of redshift  \citep{canameras15,planck26,planck39}.  The \Planck\ Collaboration predicted that \Planck's detectors would be sensitive to nearby galaxies as well as extremely high redshift galaxies, including SMGs that exhibit thermal dust emission reaching upwards to 100\% in the far-infrared \citep[FIR; see][]{efstathiou00,johnson13}.  The high redshift SMGs have a local analog in the ultra-luminous infrared galaxies (ULIRGs). An intended cross-correlation between \Herschel\ and \Planck's highest frequency channels has been acutely beneficial in discovering some of the most luminous IR galaxies, and recent \Herschel\ observations using SPIRE instrument have revealed extremely luminous dusty star forming galaxies emitting in the IR \citep[e.g.,][]{casey12,casey14,heinis13,ivison13}.  Covering the wavelengths up to 350 \micron, the \Planck\ survey is adept at detecting these highly luminous point sources across the entire sky; and, it is the first all-sky survey with the required sensitivity to systematically identify the most luminous sources in the Universe. 
			
	One of the key attributes that makes submillimetre surveys such as by \Planck\ particularly powerful is the negative k-correction in the Rayleigh-Jeans (RJ) part of the thermal dust emission that uniquely allows a nearly distance-independent study of high redshift dusty IR sources as the relative flux density observed in the submillimetre wavelengths (200-1000 \micron) is unaffected by redshift out to $z\ge10$ \citep{blain98}. This provides us with a remarkable ability to identify and investigate the frequency and the overall importance of these dusty starburst galaxies  throughout the cosmic history of galaxy assembly. Within the past decade it has also become possible to detect the CO emission lines from these galaxies using ground-based telescopes and interferometers such as the Large Millimeter Telescope Alfonso Serrano (LMT), the Jansky Very Large Array (JVLA), and the Atacama Large Millimeter Array (ALMA).  
	In addition to yielding secure spectroscopic redshifts for these optically faint population \citep[e.g.,][]{yun12,yun15,zavala15}, sensitive observations of CO emission in these galaxies can provide critical information on the total cold gas contents and their relation to the stellar and dark matter masses of the host galaxies and spatial distribution and dynamics of the ongoing star formation activities.  
		
	Given the large beam size ($\theta\ge5\arcmin$) and the high confusion noise ($>$100 mJy in the 857 GHz band), \Planck-identified high redshift sources are likely exceptional objects, either strongly lensed \citep{canameras15,geach15} or highly clustered \citep{negrello05,fu12,clements14}.  Observational data on strongly lensed objects usually require a detailed lensing model, requiring additional work to interpret the results.  On the other hand, the magnifying property of a gravitational lens can  reveal structural details of gas and star formation down to scales of 10 to 100 parsec, well beyond the reach of even ALMA or JVLA \citep[e.g.,][]{yun07,swinbank11,tamura15}.  Lensing can also allow us to probe an intrinsically much fainter population \citep{knudsen08,knudsen09,dessauges15}.  One key motivation for this study is to identify strongly lensed SMGs from the \Planck\ survey in order to exploit gravitational lensing and to probe physical details of star formation and gas properties that are not accessible through the conventional means.
	
	Here, we present our identification of eight extremely luminous infrared galaxies successfully identified as intrinsically bright and likely lensed sources through a cross-correlation analysis between the archival \Planck\ and \Herschel\ database.  This is a pilot study for a larger program to identify a large sample of extremely luminous high redshift SMGs identified by the \Planck\ survey.	In \S~2 we discuss our selection methods for the gravitationally lensed SMG candidate sources. In \S~3 we will discuss our observations of these \Planck-\Herschel\ sources using the LMT and it's two instruments: Redshift Search Receiver (RSR) and AzTEC. In \S~4 we present spectral energy distribution analysis of all of the target sources, and the detailed discussions of individual sources are summarized in \S~5. Finally, the nature of these \Planck-\Herschel\ sources are discussed in the context of the gas and dust properties derived from the new LMT observations in \S~6. We adopt a $\Lambda$CDM cosmology with $H_{0} = 70$ km s$^{-1}$ Mpc$^{-1}$ with $\Omega _{m} = 0.3$, and $\Omega _{\Lambda} = 0.7$ throughout this paper.
		
\section{Identification of Bright \Planck-\Herschel\ SMG Candidates}

	Our candidate \Planck\ SMGs were selected primarily by cross-correlating the \Planck\ Catalog of Compact Sources (PCCS), a dataset including roughly 24,000 point source objects  \citep{planck13}, with the combined catalogs of three \Herschel\ large area surveys: \Herschel\ Multi-tiered Extragalactic Survey \citep[HerMES,][]{oliver12}, \Herschel\ Stripe 82 Survey \citep[HerS-82,][]{viero14}, and \Herschel\ DDT ``Must-Do" Programmes.	These \Herschel\ source catalogs were generated using the public domain pipeline data products obtained from the \Herschel\ Science Archive (HSA)\footnote{http://www.cosmos.esa.int/web/herschel/science-archive}.  	We limited our searches only to the PCCS sources at Galactic latitude $|b| > 30\degree$ in order to minimize the Galactic source contamination. 
	
	To begin, we identified all \Herschel\ catalog sources located with a 150\arcsec\ search radius (approximately the beam size of the Planck 857 GHz band) of every PCCS source catalog position. 	Low redshift galaxies can be eliminated using the SPIRE colour \citep[see ][]{wardlow13}.  \citet{negrello10} and Wardlow et al. have reported a high efficiency (100\% and 70\%, respectively) of identifying lensed high redshift IR galaxy using $S_{500\mu} > 100$ mJy as a selection criteria. We adopt $S_{350\mu} \ge 100$ mJy as our selection method for the best candidate \Herschel\ counterpart since our sample selection relies mostly on the \Planck\ 857 GHz (350 \micron) band detection. All counterparts identified with known blazars are also removed (see below). As seen in Figure~\ref{fig:offset}, our final selected \Herschel\ counterpart candidates lie within about 50\arcsec\ ($1\sigma$) of the \Planck\ source position. The presence of a point source with this flux density indicates a rare and truly luminous and dusty source given the thermal dust emission seen in typical spectral energy distribution models \citep[$L_{IR}=1.3\times10^{13}L_\odot$ for a $S_{300\mu} =100$ mJy source at $z=2$; e.g.,][]{efstathiou00,casey12}.  Through examining the models for the expected spectral energy distribution, sources peaking at 350 \micron\ reveal a galaxy/galaxies at a high redshift ($1<z<4$), whereby the UV-heated dust is redshifted from the infrared and strongly observed in the submillimetre to millimetre wavelengths. One good indicator of a high redshift luminous dusty galaxy is the characteristic flat spectral energy distribution in the submillimetre regime of the spectrum, with the dust peak occurring within the \Herschel\ 250, 350, and 500 \micron\ band.  We note that the \Planck\ 350 \micron\ flux density is in some cases significantly larger than the \Herschel\ 350  \micron\ flux (see Table~\ref{tab:smm}), partly because \Planck\ confusion noise is high.
	
\begin{figure}
\includegraphics[width=8cm]{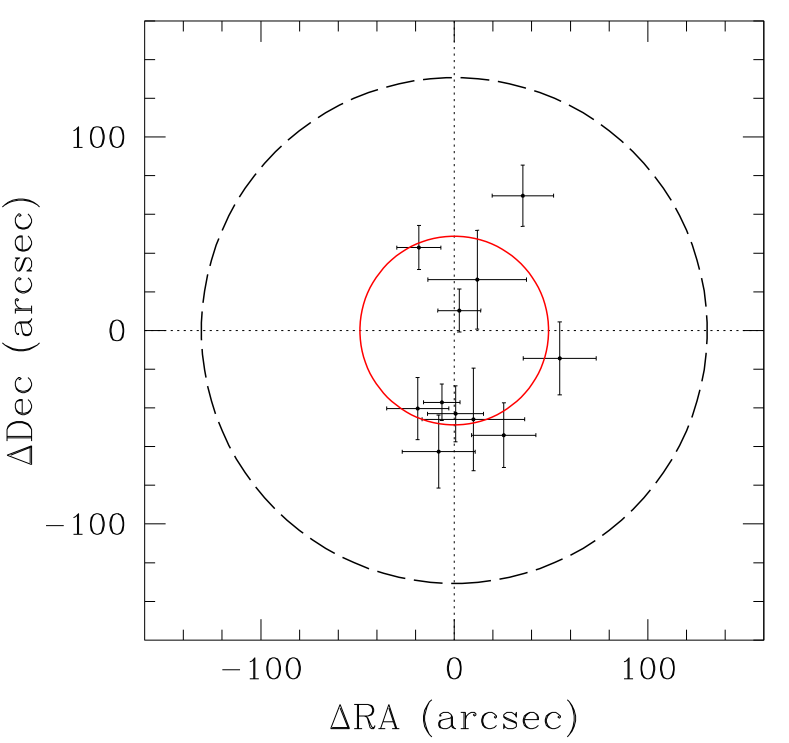}
\caption{Positional offset between the \Planck\ PCCS coordinates and the \Herschel/SPIRE 350 \micron\ source positions.  Error bars reflect the formal positional uncertainty in the individual Planck source positions, derived from the average beam size at 857 GHz (larger circle in a dashed line) and the S/N ratio.  The solid circle shows the $1\sigma$ positional offset, which is approximately $50\arcsec$ in radius.}
\label{fig:offset}
\end{figure}

	Given the large \Planck\ beam size, we use the positions of the \Herschel\ SPIRE 350 \micron\ counterpart for identifying counterparts in other wavelengths.  We exclude radio-loud AGNs and bright, late-type foreground galaxies ($z<1$) by first examining a 30\arcsec\ search radius using NASA/IPAC Extragalactic Database (NED) and the all-sky point source catalog from the Wide Field Infrared Survey Explorer (\WISE),  the NRAO VLA Sky Survey \citep[NVSS;][]{condon98}, and Faint Images of the Radio Sky at Twenty-Centimeters \citep[FIRST;][]{becker95,white97}.  We obtained images of the majority of our sources using the Sloan Digital Sky Survey (SDSS) and the Second Palomar Observatory Sky Survey (POSS-II) database to look for any possible low redshift foreground galaxies missed by other methods, and extremely faint optical counterparts are also identified in some cases.  We use \WISE\ photometry to examine the apparent brightness of these sources in the 3.4, 4.6, 11, and 22 \micron\ bands (see Table~\ref{tab:OIR}). If a source exhibits a strong detection in the 11 \micron\ and 22 \micron\ channels, a possibility of a power-law AGN is further explored.  
	
\subsection{PCCS 857 GHz and HerMES}
	The HerMES \citep{oliver12} is one of the large \Herschel\ cosmology surveys spanning a total of 94.8 deg$^2$ in combined area. We constructed a master catalog of 139,270 sources from the SPIRE photometry taken from 9 independent fields. Overall there are 19 \Planck\ point sources falling within these survey areas, and 250 matched \Herschel\ sources are identified with them.  For each of the 19 \Planck\ sources there are on average 13 \Herschel\ sources found within our 2.5\arcmin\ search radius. We identified 15 of these \Planck\ objects having at least one \Herschel\ SPIRE detection with a 350 \micron\ flux density at or greater than 100 mJy.
	
	Using NED we found that 13 of the \Planck\ candidate sources with a bright \Herschel\ counterpart are associated with low redshift galaxies or galaxy pairs.  	The remaining two \Planck\ candidates are likely to be luminous high redshift sources due to their exceedingly high flux, flat SED across the three SPIRE bands, and the absence of any bright optical counterparts in the SDSS or POSS-II plate images. One of which is found in the Akari Deep Field, with a flux density greater than 300 mJy in the SPIRE 350 \micron\ band. Because of its declination being out of reach for the LMT, this source is excluded from our follow up investigation. The last remaining source has been identified as a rediscovery of MIPS~J142824.0+352619 ($z=1.325$), also identified by \citet{wardlow13} as a gravitationally lensed candidate. Further information on this source is found in \S~5.  We note that \citet{clements14} found proto-cluster candidates in the same fields by cross-correlating the PCCS with the \Herschel\ catalogues.
	
\subsection{PCCS 857 GHz and \Herschel-Stripe 82 Survey}
	\citet{viero14} released \Herschel\ band-merged catalog of 27,257 point sources in the HerS-82 survey. Covering 79 deg$^2$, this is another major \Herschel\ wide area survey. It had been predicted by Viero et al. that the HerS-82 survey has the ability to detect roughly 100 IR galaxies at high redshift (i.e. $L_{IR}>10^{12}L_\odot$ at $z\ge2$). 	There are 42 \Herschel\ catalog matches to 16 \Planck\ candidate sources found in this survey area, and on average there are 2 to 3 \Herschel\ sources found within our 2.5\arcmin\ search radius. Of these original 16 \Planck\ sources we discover 7 objects with a \Herschel\ counterpart with $S_{350\mu} \ge 100$ mJy. Of these 7 sources, \Herschel\ data show that two are bright, low-z galaxies: CGCG~385-062 ($z = 0.047$ with $S_{350\mu} > 450$ mJy) and NGC~585 ($z = 0.018$ with $S_{350\mu} > 300$ mJy). One well-known flat spectrum radio quasar, LBQS~0106+0119 ($z=2.099$), is also rediscovered with its red SPIRE colour.   The four remaining sources are included in our list of high redshift SMG candidates. One of these sources has no NED match, and two are associated with a faint SDSS source, while the final source is associated with a 2MASS source.  Because of scheduling restrictions, we were able to conduct follow-up observations for only one (PJ020941.3) of these sources using the LMT, and only this source is considered part of our \Planck\ sample presented in this study.
	
\subsection{PCCS 857 GHz and \Herschel\ DDT ``Must-Do" Survey}
The \Herschel\ DDT ``Must-Do" Survey\footnote{http://herschel.esac.esa.int/MustDo\_Programmes.shtml}  is a targeted imaging survey of 199 PCCS sources using the \Herschel-SPIRE instrument,  designed to investigate the nature of these \Planck\ sources. We identify a total of 75 \Herschel\ sources matched to 21 \Planck\ sources meeting our selection criteria.  On average there are 3 to 4 \Herschel\ sources within the search area for each of the \Planck\ sources, and we identified a total of 13 \Herschel\ sources with $S_{350\mu} > 100$ mJy. We further excluded six well-known Quasi-Stellar Objects (QSOs) whose submillimetre emission is not associated with dust: PKS~2252-282, PKS~0402-362, ESO~362-G021, PKS~0537-441, 3C~279, and PKS~1335-127. One \Planck\ source is the \Herschel\ resolved bright, low-z galaxy, NGC~5777 ($z = 0.007$ with $S_{350\mu} > 650$ mJy). The remaining six \Planck-\Herschel\ counterparts are included in our selection of targets likely to be high redshift SMGs. 

Unknown to us initially, the \Planck\ instrument team has its own program to identify and study high redshift galaxies in their survey, and this \Herschel\ DDT ``Must-Do" Survey is apparently part of their effort to verify the candidate high redshift \Planck\ sources using higher resolution images \citep{canameras15}.  Five sources (PJ105353.0, PJ112714.5, PJ1202007.6, PJ132302.9, PJ160917.8) are in common between {Ca{\~n}ameras} et al. sample and our sample.  Detailed comparisons of individual objects are given in \S~5.

\begin{table*}
 \caption{Summary of LMT Observations}
 \label{tab:observations}
 \begin{tabular}{@{}lccccccc}
  \hline
  ID & RA & DEC & \multicolumn{2}{c}{RSR} & \multicolumn{2}{c}{AzTEC} \\
  & (J2000) & (J2000) & Dates & Int. Time (mins) & Dates & Int. Time (mins) \\
  \hline
  PJ020941.3 & 02h09m41.3s  & +00d15m59s & 1/19/14 & 60 & 1/24/14 & 10 \\
  PJ105353.0 & 10h53m53.0s  & +05d56m21s & 1/12/14 & 30 & 1/12/14 & 3.0 \\
  PJ112714.5 & 11h27m14.5s  & +42d28m25s & 1/31/14 & 60 & 5/17/14, 5/19/14 & 5.0, 10 \\
  PJ120207.6 & 12h02m07.6s  & +53d34m39s & 2/3/14 & 30 & 2/7/14 & 10 \\
  PJ132302.9 & 13h23m02.9s  & +55d36m01s & 3/21/14 & 20 & 5/10,14, 5/17/14, 5/19/14 & 10, 10, 10\\
  PJ142823.9 & 14h28m23.9s  & +35d26m20s & 3/21/14 & 30 & 6/14/14 & 20 \\
  PJ160722.6 & 16h07m22.6s  & +73d47m03s & 3/21/14, 4/29/14 & 60 & 5/10/14, 5/19/14 & 10, 40 \\
  PJ160917.8 & 16h09m17.8s  & +60d45m20s & 3/21/14 & 20 & 3/22/14, 5/17/14 & 5.0, 10 \\       
    \hline
 \end{tabular}
 \end{table*}
 
\section{LMT Observations}
	Here we present our observations of the bright \Planck-\Herschel\ sources using the LMT \citep{hughes10} atop the 15,000 ft high Volcan Sierra Negra in Puebla, Mexico. 	During the Early Science (ES) campaign when our observations were taken, the LMT was operating with a primary reflector consisting of 3 rings of fully functioning active surface system, totaling 32.5-m in diameter. 	Two facility instruments were available: the Redshift Search Receiver \citep[a broadband spectrometer system covering 72-111 GHz with a 38 GHz in total bandwidth,][]{erickson07} and AzTEC \citep[1.1mm imaging camera with 144 pixels,][]{wilson08}.  All eight \Planck-\Herschel\ sources presented here are observed with both instruments.

\begin{figure*}
\includegraphics[width=8.5cm]{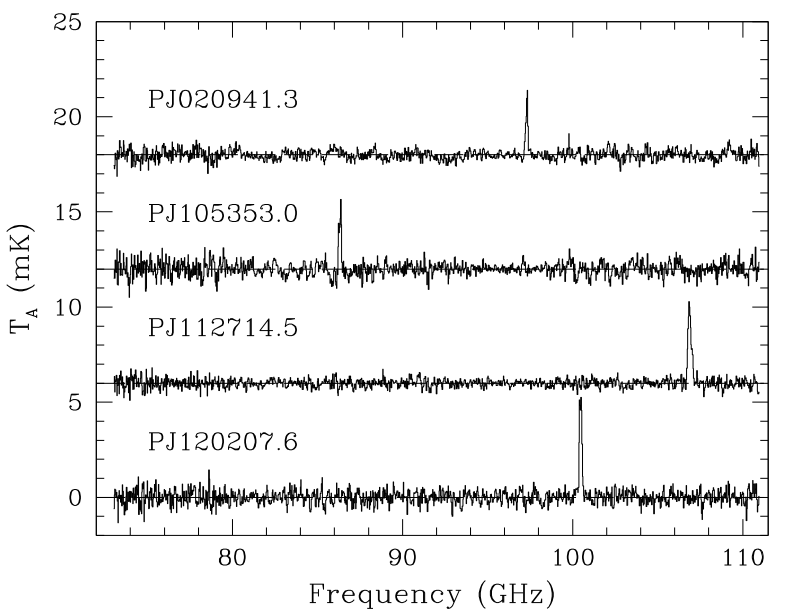}
\includegraphics[width=8.5cm]{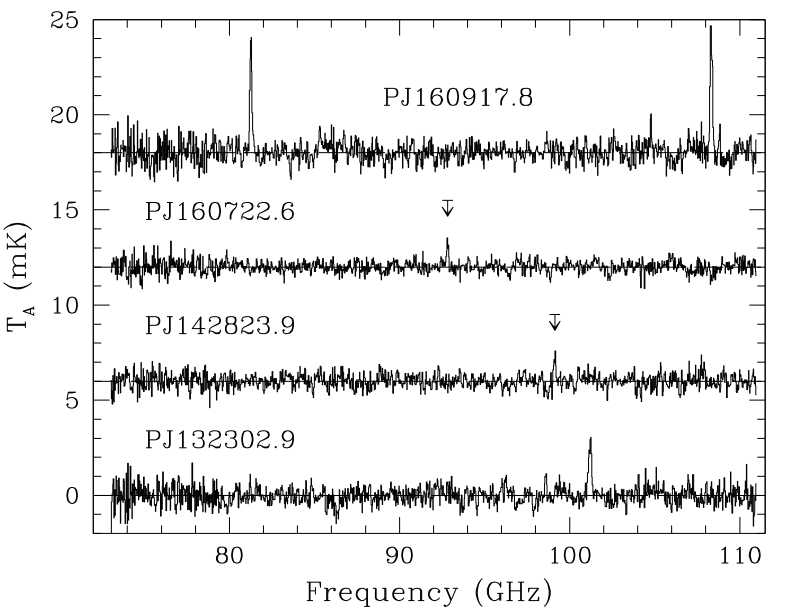}
\caption{The RSR spectra of the eight \Planck-\Herschel\ sources.  For the two sources that are detected with $S/N\sim5$ (PJ142823.9 \& PJ160722.6), the location of the CO line is shown with an arrow to guide the eye.  A more detailed zoom-in view is shown in Fig.~\ref{fig:RSRzoom}.}
\label{fig:RSR}
\end{figure*}

\begin{figure}
\includegraphics[width=8.0cm]{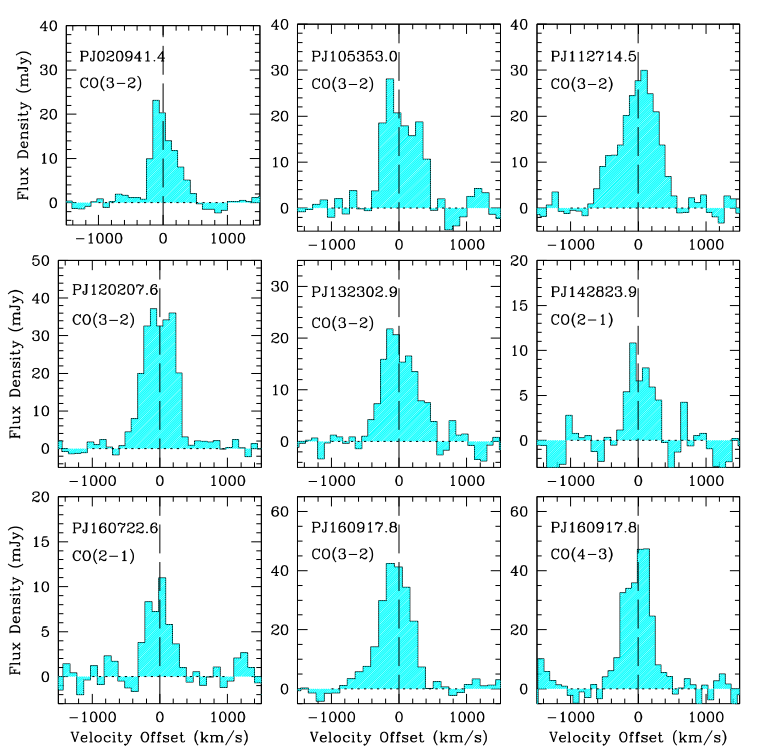}
\caption{A zoom-in view of the individual CO lines.  The velocity offset is with respect to the central redshift reported in Table~\ref{tab:RSR}.}
\label{fig:RSRzoom}
\end{figure}

\begin{table*}
 \caption{Summary of RSR Measurements}
 \label{tab:RSR}
 \begin{tabular}{@{}lccccccc}
  \hline
   ID & $\nu_{CO}$  & Line & $z_{CO}$ & $\Delta V$  & $S\Delta V$ & $\mu L'_{CO}$ & $\mu M_{H2}$ \\
   & (GHz) & & & (km/s) & (Jy km/s) & ($10^{10}$ K km/s pc$^2$) & ($10^{10} M_\odot$)  \\
 \hline
  PJ020941.3 & 97.314  & CO (3--2) & $2.5534\pm0.0002$ & $409\pm13$ & $9.5\pm0.6$ & $32\pm 2$ & $259\pm 16$  \\
  PJ105353.0 & 86.334  & CO (3--2)  & $3.0053\pm0.0003$ & $606\pm13$ & $14.1\pm0.9$ & $62\pm 4$ & $405\pm 27$  \\
  PJ112714.5 & 106.870  & CO (3--2) & $2.2357\pm0.0002$ & $652\pm13$ & $19.3\pm0.8$ & $51\pm 2$ & $335\pm 16$   \\
  PJ120207.6 & 100.471  & CO (3--2) & $2.4417\pm0.0002$ & $521\pm8$ & $21.3\pm1.0$ &  $62\pm 3$ & $405\pm 22$   \\
  PJ132302.9 & 101.208  & CO (3--2) & $2.4167\pm0.0004$ & $601\pm14$ & $13.3\pm1.1$ & $40\pm 3$ & $265\pm 22$   \\
  PJ142823.9 & 99.125  & CO (2--1) & $1.3257\pm0.0003$ & $422\pm21$ & $4.5\pm0.7$ & $10\pm 2$ & $54\pm 11$   \\
  PJ160722.6 & 92.883  & CO (2--1) & $1.4820\pm0.0004$ & $372\pm23$ & $4.0\pm0.6$ & $11\pm 2$ & $59\pm 11$  \\
  PJ160917.8 & 81.262  & CO (3--2) & $3.2553\pm0.0003$ & $504\pm9$ & $23.7\pm1.6$ & $119\pm 8$ & $778\pm54$  \\       
                     & 108.326  & CO (4--3) & $3.2561\pm0.0003$ & $450\pm6$ & $21.1\pm1.4$ & $60\pm 4$ & $562\pm 38$  \\       
 \hline
 \end{tabular}
 \medskip
 
(1) A conversion of 7 Jy/K is adopted to convert the measured antenna temperature in $T_A*$ to flux density, using the calibration factor derived between December 2013 and January 2014.

(2) Unknown lensing amplification $\mu$ is reflected in the derived CO luminosity and \H2\ mass as $\mu L'_{CO}$ \& $\mu M_{H2}$.

 \end{table*}

\subsection{Redshift Search Receiver Measurements}
	Our observations include redshift measurements taken with the Redshift Search Receiver operating in the frequency window of 73--111 GHz, with 4 detector pixels organized in a dual--beam, dual--polarization configuration. The RSR beam-switches at 1 kHz between the two beams separated by 78$\arcsec$ in Azimuth direction on the sky, providing a stable baseline. The backend receiver consists of an auto-correlator system that covers a total of 38 GHz spectral bandwidth simultaneously with a spectral resolution of 31.25 MHz.	Table~\ref{tab:observations} provides information of the integration time for each source. For the majority of the sources a CO line was evident within 15 minutes of integration, and only 3 of our sources needed a longer integration to yield a CO detection with a $S/N$ ratio $\ge4$. 
	
The measured RSR spectra of all eight observed sources covering the entire redshift range are shown in Figure~\ref{fig:RSR}, and the measured and derived CO line quantities are summarized in Table~\ref{tab:RSR}.  As discussed in detail by \citet{yun15}, two or more CO transitions fall within the RSR spectral band at $z\ge3.15$, and a unique redshift for the CO source can be determined from the RSR spectrum alone (as is the case for the $z=3.256$ source PJ160917.8).  If only one CO line is detected as in the seven out of eight cases shown in Fig.~\ref{fig:RSR}, additional information is required to determine the exact redshift.  We derive a unique redshift for all eight \Planck-\Herschel\ sources observed using their panchromatic photometric redshift support as discussed in Appendix~\ref{sec:appendixRSR}.

The zoom-in view of the individual CO lines in Figure~\ref{fig:RSRzoom} clearly shows that all of the CO lines are fully resolved by the RSR, some with the characteristic ``top hat" profile commonly seen for CO lines in rotationally supported galaxies, while others show an asymmetric profile of a more complex kinematics, perhaps resulting from differential lensing.  The measured line widths $\Delta V$, corrected for the instrumental resolution (see appendix), ranges between 372 and 652 \kms.   For the five sources in common with \citet{canameras15}, the measured line widths and line integrals agree well with theirs, with the median ratios of 1.1 and 0.81, respectively, except for PJ105353.0 (our line width and line flux are factors of 1.6 and 1.9 larger).

The CO line luminosity, $L'_{CO}$, is computed for each observed CO transition using Eq.~(3) by \citet{solomon97},
$$L'_{CO} = 3.25\times10^7 S_{CO}\Delta V \nu_{obs}^{-2} D_L^2(1+z)^{-3},$$ with $S_{CO}\Delta V$ in Jy km s$^{-1}$, $\nu_{obs}$ in GHz, and $D_L$ in Mpc.  The molecular gas mass $M_{H2}$ is derived first by converting these line luminosities to $L'_{CO(1-0)}$ using the average ``SMG" ratios of $L'_{CO(2-1)}/L'_{CO(1-0)}=0.85$, $L'_{CO(3-2)}/L'_{CO(1-0)}=0.66$, and $L'_{CO(4-3)}/L'_{CO(1-0)}=0.46$ \citep{carilli13} and then applying the standard CO-to-\H2\ conversion factor of $\alpha_{CO}=4.3\,M_\odot$ [K \kms\ pc$^2$]$^{-1}$ \citep{bolatto13}.  As discussed in detail by Bolatto et al. in their review article, this conversion from $L'_{CO(1-0)}$ to $M_{H2}$ can vary by a factor of up to 5 or more, depending on metallicity, excitation, and dynamical state of the gas.  However, there are no compelling reasons to believe any of these concerns are important for these \Planck-\Herschel\ or other high redshift sources \citep[see the detailed discussions by][]{scoville15}, and we adopt the standard Galactic conversion factor.  The derived molecular gas masses $M_{H2}=(5.5-78)\times 10^{11}\mu^{-1} M_\odot$ (uncorrected for an unknown magnification factor $\mu$) are extremely large and make these \Planck-\Herschel\ sources some of the most luminous CO emitters ever detected -- see Figure~9 by \citet{carilli13} for a comparison.  One of these sources, PJ020941.3, is already shown to be a strongly lensed object with a magnification of $\mu \approx 13$ \citep{geach15}, yielding the intrinsic gas mass of $2.0\pm0.1\times 10^{11} M_\odot$.  
Even if they are all strongly magnified with $\mu\sim 10$, their gas masses corrected for lensing are still comparable to some of the most gas-rich SMGs ever studied. 


\begin{figure}
\includegraphics[width=0.95\columnwidth]{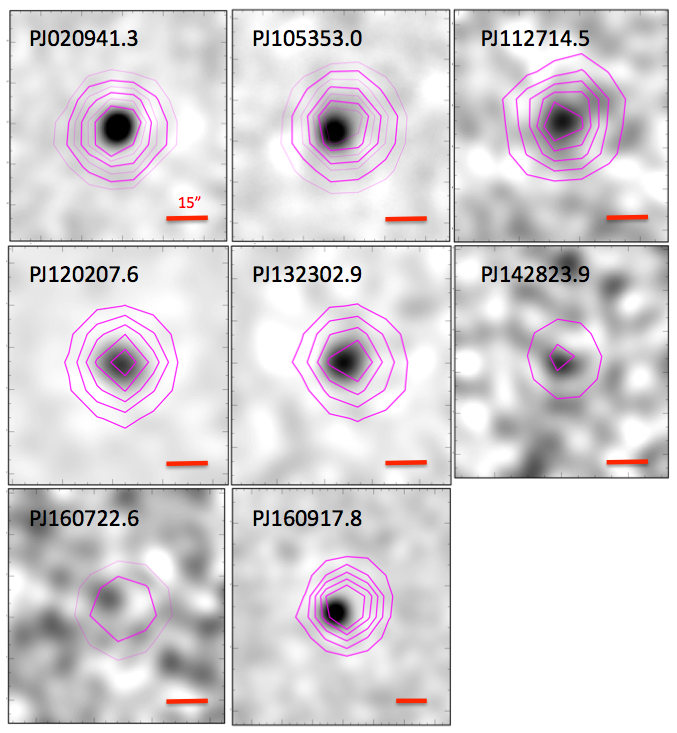}
\caption{AzTEC 1.1mm continuum images of the 8 \Planck-\Herschel\ sources.  The 1.1mm continuum is essentially unresolved at 8.5$\arcsec$ resolution of AzTEC on LMT and are mostly centered on the peak of the \Herschel\ 350 \micron\ image shown in contours (increments of 100 mJy, except for PJ160722.6, which is in increment of 50 mJy).}
\label{fig:aztec}
\end{figure}

\subsection{AzTEC Measurements}
	Continuum observations were made with the AzTEC
camera~\citep{wilson08}, a 1.1mm continuum camera that is a facility
instrument on the LMT  between March 21 and May 18 of 2014.   Only the inner 32-m diameter area of the LMT's primary mirror
was illuminated during these Early Science observations, leading to a beam diameter (FWHM) of 8.5\arcsec.  The
telescope executed a modified Lissajous pattern in order to modulate the telescope bore-sight on the sky.  
This pattern is described in \citet{wilson08} though the pattern parameters were adjusted for
the LMT to provide a fully-sampled 1.5\arcmin\ area of sky with near-uniform coverage.  For each source, the pattern was centered on
the location of the \Herschel\ source coordinates matched to the low resolution \Planck\ point source detections. The observations were
made over a wide range of atmospheric conditions with 225~GHz opacities ranging from $\tau_{225}=0.09$ to 0.3.  On-source integration times
varied from 3-40 minutes depending on the quality of the detection as measured in real time.
Pointing corrections were applied for each observation based on pointing measurements of nearby radio quasars made immediately prior
to or immediately following the science observation.  The measured pointing corrections with respect to the nominal LMT pointing model
were $1-4$\arcsec\ in azimuth and $9-20$\arcsec in elevation.

AzTEC observations were calibrated using a set of 106 beam-maps of bright radio quasars and asteroids made during the full
observing season.  The full set of maps were used to measure the elevation-dependent gain of the telescope.  
For these observations a correction as large as a factor of 1.4 in gain was required for the lowest-elevation observations. The
resulting flux uncertainty, as measured by a concurrent monitoring program is 10 per cent.  We note that, starting from the fall of 2014, the removal of an astigmatism term from the primary mirror using the active surface system resulted in a flat gain curve for the entire elevation range.  

The data for each source was reduced using the AzTEC Standard Pipeline described in detail in \citet{wilson08} and \citet{scott08}.
Since we are nominally searching for point sources, we use the standard Wiener filtering technique described in \citet{perera08}.
The pipeline produces a filtered signal map (in Jy/beam), a filtered weight map (in beam$^2$/Jy$^2$), and a filtered point spread function.
Because of the Wiener filter, the flux of any detected point source may be read directly from the peak-flux pixel near the center of the
source.  A typical noise in the AzTEC images is $\sigma \sim 1.5-4$ mJy, and our
target sources are detected with $S/N$ = 4-20 (see Fig.~\ref{fig:aztec}).  These values are reported in Table~\ref{tab:smm}, 
with a minimum measurement uncertainty of 10 per cent for sources detected with $S/N>10$.

\begin{table*}
\caption{Summary of  \Herschel\ SPIRE, \Planck, and AzTEC 1100 \micron\ photometry}
\label{tab:smm}
\begin{tabular}{@{}lccccc}
\hline
\hline
Source ID & $SPIRE_{250} $ (mJy) &   $SPIRE_{350} $ (mJy)  &  $SPIRE_{500} $ (mJy) & $Planck_{350} $ (mJy) & $S_{1100} $ (mJy) \\
\hline
PJ020941.3 & $824\pm82$ &	$897\pm89$ & $703\pm70$ & $877\pm375$ & $147\pm15$ \\
PJ105353.0 & $1028\pm 103$ & $996\pm100$ & $735\pm74$ & $2780\pm684$ & $98\pm10$  \\
PJ112714.5 & $810\pm81$ &	$708\pm71$ & $448\pm45$ & $965\pm191$ & $24.0\pm2.0$ \\
PJ120207.6 & $618\pm62$ & $646\pm65$ & $451\pm45$ & $401\pm225$ & $68.7\pm7.0$ \\
PJ132302.9  & $652\pm65$ & $590\pm59$ & $393\pm39$ & $832\pm214$ & $32.7\pm3.0$ \\
PJ142823.9 & $315\pm32$ & $239\pm24$ &  $135\pm14$ & $424\pm323$ & $14.4\pm2.0$ \\
PJ160722.6 & $167\pm17$ & $167\pm17$& $78\pm8.0$ & $888\pm270$ & $7.6\pm1.0$ \\
PJ160917.8 & $568\pm57$ & $693\pm69$ & $512\pm51$ & $1083\pm255$ & $73.8\pm7.0$ \\
\hline
 \end{tabular}
 \end{table*}

\section{Spectral Energy Distribution Analysis of the \Planck-\Herschel\ Sources}

\subsection{Modified Blackbody Model \label{sec:BB}}

\begin{figure*}
\includegraphics[width=1.4\columnwidth]{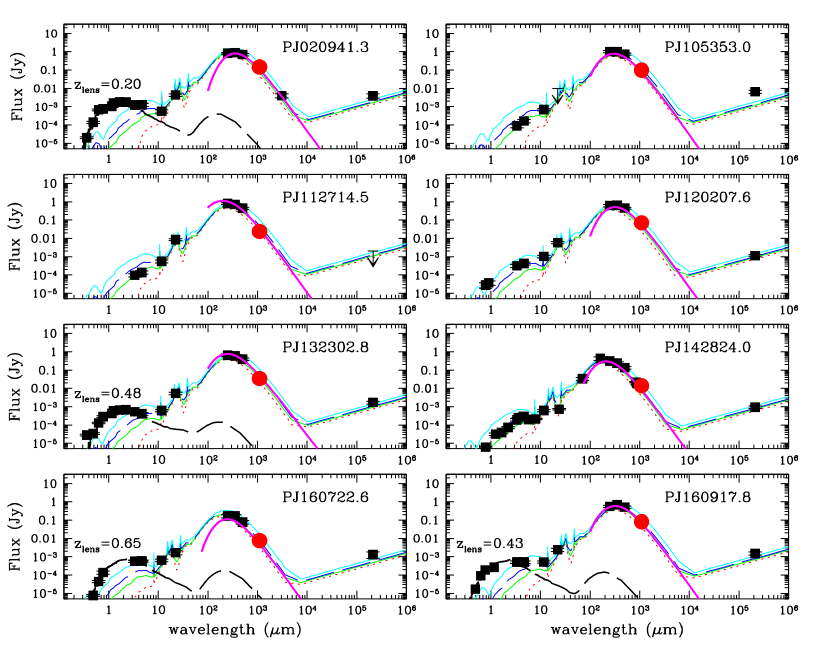}
\caption{Spectral Energy Distributions (SEDs) of the eight \Planck-\Herschel\ sources. The models shown in solid, dotted, dashed, and dot-dashed lines are the best fit dusty starburst SEDs with different starburst ages by \citet{efstathiou00} at the fixed CO redshifts (see Table~\ref{tab:RSR}).  Modified blackbody models fitting only the \Herschel\ and AzTEC photometry are shown in thick magenta lines.   In all cases, the new AzTEC 1100 \micron\ points are shown as filled circles probe the RJ portion of the dust spectra.  An elliptical galaxy SED template by \citet{polletta07} covering the optical wavelength bands is also shown using a black dashed line for the four sources with an obvious foreground galaxy in the SDSS plates. }
\label{fig:SEDs}
\end{figure*}

We explore constraints on the infrared luminosity and attempt to characterize the dust emission for these systems by analyzing their SEDs using a modified blackbody model.  We adopt the functional form based on the derivation by \citet{yun02} for the modified blackbody model.  A detailed accounting of our method is outlined in Appendix~\ref{sec:appendixMBM}.   As summarized in Table~\ref{tab:chisq}, the characteristic dust temperature $T_d$ and IR luminosity, derived primarily from the three \Herschel\ SPIRE band photometric measurements (250, 350, and 500 \micron) and the AzTEC 1100 \micron\ continuum measurements (see Table~\ref{tab:smm}), range between 43-84 K and $L_{IR}=(1-28)\times 10^{13} \mu^{-1} L_\odot$.  There is a good constraint on most sources (see Fig.~\ref{fig:modifiedBB}), although dust temperature is not constrained well for some of our sources.  The best fit models are also shown in solid magenta lines in Figure~\ref{fig:SEDs}, suggesting that the wavelength coverage of our photometric data is not sufficient to fully probe their dust SED in some cases.   The star formation rates in  Table~\ref{tab:chisq} are calculated using the empirical calibration by \citet{kennicutt98}, corrected for the Kroupa IMF [i.e., $SFR=L_{IR}/(9.4\times 10^9 L_\odot)\, M_\odot /yr$].


\subsection{Template SED Analysis \label{sec:SBSED}}

\begin{figure*}
\includegraphics[width=14.0cm]{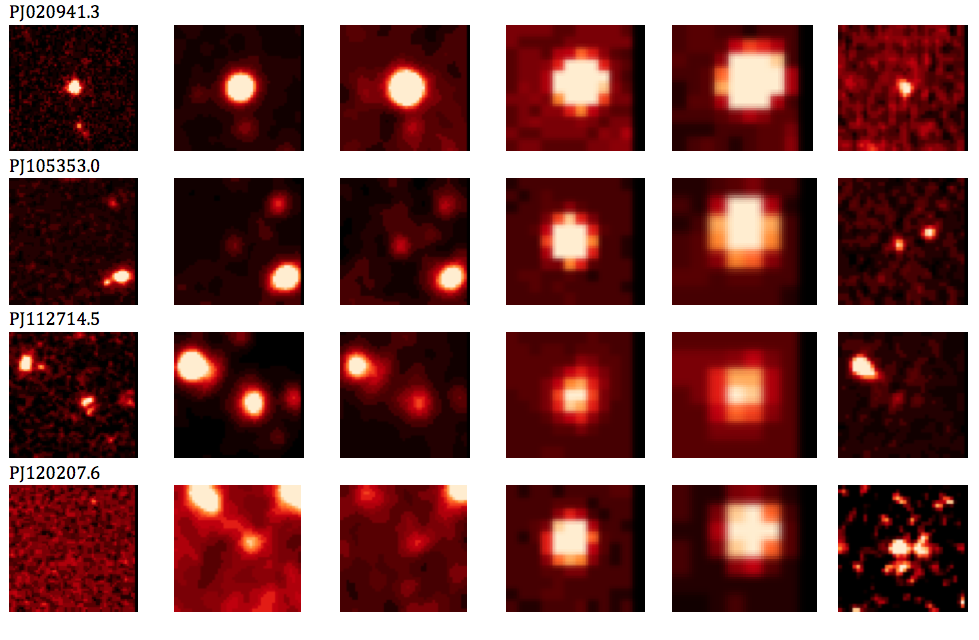}
\caption{Multi-wavelength cutout images of each of the four \Planck-\Herschel\ sources. Each $60 \arcsec $ by $60 \arcsec $ cutout corresponds to: (left to right) Optical (SDSS), \WISE\ W1(3.4$\mu$) \& W2(4.6$\mu$), \Herschel\ SPIRE 250 \micron\ \& 350 \micron, and VLA (FIRST 1.4 GHz).  }
\label{fig:postage1}
\end{figure*}

\begin{figure*}
\includegraphics[width=14.0cm]{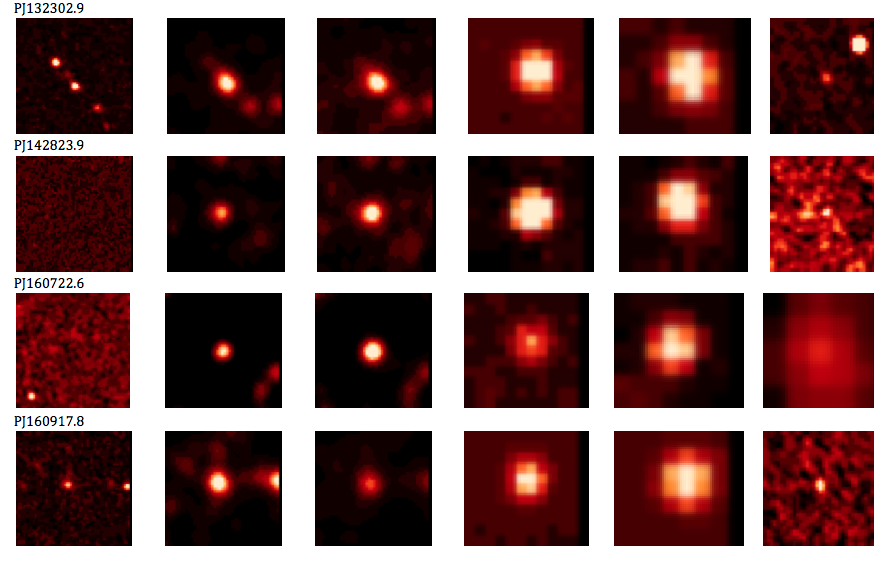}
\caption{Multi-wavelength cutout images of the remaining four \Planck-\Herschel\ sources.  The panels shown are identical to Fig.~\ref{fig:postage1}.   PJ160722.6 lies out of the coverage range for FIRST and SDSS surveys, and we show instead the NVSS 1.4 GHz and 2MASS $K_{s}$ band images. }
\label{fig:postage2}
\end{figure*}

In addition to their \Herschel\ SPIRE and AzTEC photometry that trace the peak of the dust SED, all eight \Planck-\Herschel\ sources are clearly detected in a wide range of wavelength bands, from optical to radio as shown in Figures \ref{fig:postage1} \& \ref{fig:postage2} (also see Table~\ref{tab:OIR}).  By analysing these SEDs using model templates, it may be possible to gain additional insights into the nature of their extreme luminosity.   In particular, we compare the observed SEDs with the model SEDs of young stellar clusters embedded in a giant molecular cloud and evolved by different amounts of time by \citet{efstathiou00} in Figure~\ref{fig:SEDs}. Since the discrete SED templates available cannot be tuned for a particular object, especially in the UV, optical, and near-IR where source and dust geometry can lead to a large dispersion, we only fit the long wavelength data at $\lambda\ge70$ \micron. The good agreement between the data and the models suggests that intense star formation activities can account for most of the observed properties in nearly all cases.    The radio to millimeter wavelengths part of the SED is constructed using the well established radio-IR correlation for star forming galaxies as described by \citet{yun02}, and the observed 1.4 GHz radio emission is also consistent with the radio and IR luminosity being powered by a pure starburst in nearly all cases -- see additional discussions on the nature of the powering source in \S~\ref{sec:SBAGN}.


Using the best fit SED models shown in Figure~\ref{fig:SEDs}, we compute the template based IR luminosity ($L_{IR,SB}$) between 8-1000 \micron\ and  star formation rate $SFR_{SB}$ for each \Planck-\Herschel\ source (see Table~\ref{tab:chisq}).  The IR luminosity derived from the best fit template SED ranges between $(1.4-29)\times 10^{13} \mu^{-1} L_\odot$, in good agreement with the estimates from the modified blackbody model calculation discussed above (\S~\ref{sec:BB}). This is not surprising given that the \Herschel\ SPIRE photometry points constrain the bulk of the luminosity in both models.    

A significant deviation from the Efstathiou templates is seen in the rest frame optical portion of the SEDs in at least four cases.  A subset of these sources are already known to be lensed by a foreground galaxy \citep[see \S~\ref{sec:IND} and][]{canameras15}, and the presence of one or more foreground lensing galaxy along the same line of sight is naturally expected.  Indeed, for PJ020941.3, PJ132302.8, PJ160722.6, and PJ160917.8, an addition of an elliptical galaxy SED by \citet{polletta07} nicely accounts for the excess flux in the rest frame optical bands, as shown in Figure~\ref{fig:SEDs}.  This photometric evidence for a foreground lensing galaxy is further discussed below for individual objects (see \S~\ref{sec:IND}).

\begin{table*}
\caption{Best Fit Parameters of the Starburst SED Template and modified blackbody models for each source with derived dust and ISM masses}
\label{tab:chisq}
\begin{tabular}{@{}lccccccc}
\hline
Source ID & $\mu L_{IR,SB}^\dagger$ & $\mu SFR_{SB}$ & $T_{d} $    & $\mu L_{IR,BB}$ & $\mu SFR_{BB}$   & $\mu M_{d}$ & $\mu M_{ISM} $ \\
		& ($10^{14}  L_\odot)$ & ($M_\odot$ yr$^{-1}$)  & (K) & ($10^{14}  L_\odot)$ &	($M_\odot$ yr$^{-1}$)  &  ($10^{10} M_\odot$)  & ($10^{10} M_\odot$) \\		
\hline
PJ020941.3 & $2.0\pm0.4$ & 21,276 & $42.6_{-4.8}^{+6.4}$ & $ 1.4 \pm 0.07 $ & 15,027 & $2.16 \pm 0.43 $ & $512 \pm 102$ \\ 	
PJ105353.0 & $2.9\pm0.4$ & 30,681 & $62.4_{-7.6}^{+12.6}$ & $ 2.8 \pm 0.13 $ & 29,983  & $ 1.33 \pm 0.27  $ & $ 257 \pm 51 $ \\	
PJ112714.5 & $1.1\pm0.2$ & 12,021  & $ 84.0_{-15.6}^{+30.4}$ & $ 2.24 \pm 0.28 $ & 23,816  & $ 0.37  \pm  0.07 $ &  $ 102 \pm 20 $ \\	
PJ120207.6 & $1.4\pm0.3$ & 15,000  & $47.2_{-5.6}^{+7.0}$ & $ 1.0 \pm 0.13 $ & 10,639  & $ 1.03 \pm 0.21 $ & $ 257 \pm 51 $ \\	
PJ132302.9  & $1.2\pm0.2$ & 13,085 & $64.4_{-7.8}^{+16}$ & $ 1.26 \pm 0.18 $ & 13,393  & $ 0.50 \pm 0.10 $ & $ 124 \pm 25 $ \\	
PJ142823.9 & $0.19\pm0.04$ & 2,138  & $48.8_{-5.2}^{+7.2}$ & $ 0.20 \pm 0.05 $ & 2,123  & $ 0.25 \pm 0.05 $ & $ 101 \pm 20 $ \\	
PJ160722.6 & $0.14\pm0.03$ & 1,503  & $42.6_{-6.2}^{+6.8}$ &$ 0.10 \pm 0.12 $& 1,064 & $ 0.13 \pm 0.03 $ & $ 50 \pm 10 $ \\		
PJ160917.8 & $2.0\pm0.4$ & 21,226  & $62.6_ {-6.8}^{+8.6}$ &$ 2.0 \pm 0.1 $ & 21,226 & $ 0.96 \pm 0.20 $ & $ 165 \pm 33 $ \\	
\hline
\end{tabular}

\medskip
    $^\dagger$  $L_{IR}$ is the far-infrared luminosity derived by integrating between 8-1000 \micron\ in wavelength \citep{sanders96}. The exceedingly high dust masses are likely magnified by an unknown magnification factor $\mu$ due to lensing. 
\end{table*}

 \begin{table*}
\caption{Summary of Optical and Near-IR photometry}
\label{tab:OIR}
\begin{tabular}{lccccccc}
\hline
\hline
Source ID & $J_{1.26\mu}$ (mJy)  & $H_{1.6\mu}$ (mJy) & $ K_{2.22\mu}$ (mJy)  & $W_{3.4\mu}$  (mJy) & $W_{4.6\mu} $  (mJy) & $W_{11\mu}$  (mJy) & $W_{22\mu}$ (mJy) \\
\hline
PJ020941.3 & $0.04\pm0.00*$ & $0.06\pm0.01*$ & $0.05\pm0.01*$ & $1.22\pm0.03$ & $1.30\pm0.03$ & $0.61\pm0.12$ & $4.35\pm0.87$ \\
PJ105353.0 & $0.24\pm0.00*$ & $0.27\pm0.00*$ & $0.22\pm0.01*$ & $0.09\pm0.01$ & $0.17\pm0.02$ & $0.70\pm0.15$ & ... \\
PJ112714.5 & $0.23\pm0.04$ & $0.27\pm0.06$ & $0.31\pm0.07$ & $0.10\pm0.01$ & $0.14\pm0.01$ & $0.57\pm0.13$ & $8.65\pm0.90$ \\
PJ120207.6 & ... & ... & ... & $0.32\pm0.01$ & $0.44\pm0.02$ & $1.04\pm0.10$ & $5.89\pm0.80$ \\
PJ132302.9 & $0.61\pm0.04$ & $0.67\pm0.06$ & $0.68\pm0.06$ & $0.55\pm0.01$ & $0.42\pm0.02$ & $0.63\pm0.10$ & $5.45\pm0.78$ \\
PJ142823.9 & ... & ... & ... & $0.19\pm0.01$ &$ 0.30\pm0.01$ & $0.67\pm0.09$ & ... \\
PJ160722.6 & ... & ... & ... & $0.58\pm0.01$ & $0.60\pm0.01$ & $0.65\pm0.06$ & $1.70\pm0.47$ \\
PJ160917.8 & $0.05\pm15.72$ & $0.89\pm0.09$ & $0.68\pm0.11$ & $0.53\pm0.01$ & $0.52\pm0.01$ & $0.52\pm0.05$ & $2.43\pm0.42$ \\
\hline
\end{tabular}

\medskip
      (1) For JHK Photometry, a star(*) indicates a UKIDSS photometry while the remaining photometry come from the 2MASS catalog.
\end{table*}
 
\section{Analysis of Individual Sources}
\label{sec:IND}

\subsection{PJ020941.3}

PJ020941.3 is a \Planck\ source found in the  \Herschel-Stripe 82 Survey field and is associated with the $z=0.202$ galaxy SDSS~J020941.27+001558.5.    Our photometric redshift analysis predicts a lensed, dusty starburst galaxy at $z\approx2.5$ with the dust peak near 350 \micron\ with $S_{350\mu} = 897\pm89$ mJy.  We detect 1100 \micron\ continuum emission easily using AzTEC, measuring $S_{1100\mu} = 147 \pm 20 $ mJy, centered on the \Herschel\ source position (see Fig.~\ref{fig:aztec}). Using the RSR we detected CO (3-2) line at 97.31 GHz, yielding a redshift of $z = 2.5536\pm0.0002$ (a single line redshift resolved using the photometric redshift support as shown in Fig.~\ref{fig:photoz}). 


The nature of this source, also known as `9io9' \citep{geach15}, is unique due to the fact that two unrelated research teams have identified it entirely independently. Our initial selection of PJ020941.3 arose from matching \Planck's 857 GHz position coordinates to \Herschel\ SPIRE observations in the HerS-82 survey (see $\S$ 2.3). As we were ready to observe this source using the LMT, a Director's Discretionary Time request was made by an independent team based in the United Kingdom. The preliminary results from VISTA-CFHT Stripe 82 survey (Geach et al. in prep) were combined to form multiple $iJK_{s}$ band images of potential lens candidates. Tens of thousands of images were passed along to the SPACEWARPS citizen science project (Marshal et al. in prep, More et al. in prep), whereby this collaboration made its way to the British Broadcasting Corporation (BBC)  \textit{Stargazing Live!} broadcast from 7th January 2014 to 9th January 2014. Citizen scientists were trained to identify the features reflective of gravitational lensing, and the partial red Einstein ring of '9io9' was chosen as the best lens candidate.  Their follow-up spectroscopy yielded redshifted optical emission lines at $z=2.553$, and their lensing model constructed from near-infrared imaging data suggests a lensing magnification of $\mu\approx10$.

\subsection{PJ105353.0}
One of  \Planck\ sources observed by the "Must-Do" \Herschel\ SPIRE program, PJ105353.0, also known as the radio source NVSS~J105352+055623, is one of our brightest sources. It has a 350 \micron\ flux density of $996\pm100$ mJy, and it peaks in the 250 \micron\ band at $1028\pm103$ mJy.  This is the second brightest AzTEC source we detected with $S_{1100\mu}=98\pm10$ mJy, well centered on the \Herschel\ SPIRE 350 \micron\ peak position (see Fig.~\ref{fig:aztec}).  The bright line detected by the RSR at 86.334 GHz  (Fig.~\ref{fig:RSR}) is robustly identified as CO (3--2) line by the photometric redshift support (see Fig.~\ref{fig:photoz}), placing this object at $z =  3.0053\pm0.0003$.  No bright optical source is seen on any of the SDSS images. Its flat spectral energy distribution in the SPIRE bands clearly indicates that this is indeed a high redshift source, and its exceedingly high flux is expected to be a lensed thermal emission from a dusty background source. The 1.4 GHz radio continuum reported by the NVSS, $6.5\pm1.3$ mJy,  is slightly above what is expected from its far-IR emission (see Fig.~\ref{fig:SEDs}), and this ``radio-excess" may be an indication of a radio AGN \citep[see][]{yun01}.  \citet{canameras15} identify the single line detected using the IRAM 30m telescope as CO (3--2) line at ``$z=3.0$", and their 2\arcsec resolution 870 \micron\ image obtained using the Submillimeter Array (SMA) shows a single source without clear evidence for lensing.

\subsection{PJ112714.5}
PJ112714.5  is a \Herschel\ ``Must-Do" survey source and is one of our brightest sources with a flat SED across the \Herschel\ SPIRE channels, peaking in the 250 \micron\ band.  
An AzTEC source with $S_{1100\mu}=24\pm2$ mJy is detected coincident with the \Herschel\ source as seen in Fig.~\ref{fig:aztec}, and it has a red \WISE\ mid-IR counterpart with detections in all four bands.  The observed SED supports the RSR detected line at 106.877 GHz to be CO (3-2) at a $z = 2.2357\pm0.0002$ (see Fig.~\ref{fig:photoz}).   \citet{canameras15} identify the single line they detected using the IRAM 30m telescope as CO (3--2) line at ``$z=2.2$".  Their 2\arcsec resolution 870 \micron\ SMA image shows a single, slightly extended source without a clear optical counterpart, offset from a group of galaxies at $z\approx 0.3$, suggesting possible lensing by a galaxy group rather than by a single foreground galaxy \citep[e.g.,][]{george13,timmons16}.  An NVSS radio source with $S_{1.4 GHz} = 31$  mJy (see Fig.~\ref{fig:postage1}) is also found just 18 \arcsec\ away, suggesting a rather confused and busy foreground environment.


\subsection{PJ120207.6}
PJ120207 is another \Herschel\ ``Must-Do" survey source and is one of our brightest 350 \micron\ peakers ($S_{350\mu} = 646$ mJy) with $L_{IR} \ge 10^{14} L_\odot$.  The observed SED is flat across the 250 and 350 \micron\ bands and drops slightly at 500 \micron. An AzTEC source with $S_{1100\mu}=69\pm7$ mJy is detected coincident with the \Herschel\ source as seen in Fig.~\ref{fig:aztec}, and it has a red \WISE\ mid-IR counterpart with detections in all four bands.  No optical galaxy is found at the AzTEC position, but there is an SDSS source (SDSS J120207.68+533432.5) located 6.8\arcsec\ away from the AzTEC position, and this $z_{ph}\approx 0.43$ galaxy is a possible lensing source \citep{canameras15}.  The RSR has detected a single line at an observed frequency of 100.501 GHz, which we identify as CO (3--2) line at $z = 2.4417\pm0.0002$.  \citet{canameras15} identify the same line detected using the IRAM 30m telescope as CO (3--2) line at ``$z=2.4$".  Their 2\arcsec resolution 870 \micron\ SMA image shows two sources separated by  about 4\arcsec, suggesting lensing by a foreground galaxy.  No radio counterpart is found for this object in the VLA FIRST survey.  

\subsection{PJ132302.9}
PJ132302 is a bright 350 \micron\ source ($S_{35mu} = 590$ mJy) with $L_{IR} \ge 10^{14} L_\odot$, and the dust peak of this \Herschel\  ``Must-Do" survey source occurs in the 250 \micron\ band with $S_{250\mu} = 652$ mJy.  An AzTEC source with $S_{1100\mu}=33\pm3$ mJy is detected coincident with the \Herschel\ source as seen in Fig.~\ref{fig:aztec}, and it also has a red \WISE\ mid-IR counterpart with detections in all four bands (see Fig.~\ref{fig:postage2}).   The SDSS image shows a crowded field with a cluster of sources that are identified as ``stars" by the SDSS database.  The nearest match to the \WISE\ source is SDSS~J132302.83+5535555.1, which is consistent with a $z=0.48$ galaxy when analysed together with the \WISE\ and 2MASS photometry (see Fig.~\ref{fig:SEDs}), possibly representing the foreground lensing galaxy.    The RSR has detected a single line at an observed frequency of 101.202 GHz, which we identify as CO (3--2) line at $z = 2.4167\pm0.0004$.  \citet{canameras15} identify the same line detected using the IRAM 30m telescope as CO (3--2) line at ``$z=2.4$".  Their 2\arcsec resolution 870 \micron\ SMA image shows an unresolved source centered on a faint optical source about 3\arcsec northeast of the SDSS galaxy.  The 1.4 GHz FIRST source VLA~J132303.1+553558.6 with $S_{1.4GHz}=2.12\pm0.26$ mJy coincides with the SMA source position.  

\subsection{PJ142823.9}
PJ142823.9 is a bright \Herschel\ source previously identified as `HBootes03'  by \citet{wardlow13} and is also a well-known $z=1.325$ \Spitzer-identified hyperluminous infrared galaxy (HyLIRG) MIPS~J142824.0+352619 \citep{borys06,swinbank06}.  An extensive discussion of this object is found in these papers, and here we highlight only the details relevant to our discussions.  We identify the line detected at 99.131 GHz as that of CO (2--1) line at  $z = 1.3257\pm0.0003$, and our measured line integral of $S\Delta V=4.5\pm0.7$ Jy \kms\ and  the line width of $\Delta V = 422\pm21$ \kms\ agree very well with the previously reported Nobeyama Millimeter Array measurements by \citet[][$S\Delta V=5.3\pm1.2$ Jy \kms, $\Delta V = 386\pm64$ \kms]{iono06}.   \citet{borys06} have suggested gravitational lensing by a foreground $z=1.034$ as a possible explanation for the large IR luminosity, with a modest magnification factor ($\mu<10$).  \citet{hailey10} detected the 158-\micron\ [C~II] fine-structure line and derived a relatively large $L_{[CII]}/L_{FIR}$ ratio of $\approx 2\times 10^{-3}$, much higher than those found in the local ULIRGs and high-redshift QSOs and more similar to the values seen in the local starburst galaxies like M82.  This large $L_{[CII]}/L_{FIR}$ ratio would also be naturally explained by gravitational lensing.


\subsection{PJ160722.6}
PJ160722.6 is the faintest \Herschel\ source matched with the \Planck\ catalog, with $S_{250\mu} = 167$ mJy and $S_{350\mu} = 167$ mJy, and its flat SED across the \Herschel\ SPIRE bands is suggestive of a high redshift source.  Our AzTEC imaging revealed a relatively faint source coincident with the \Herschel\ source (see Fig.~\ref{fig:aztec}), with $S_{1100\mu}=7.6\pm1.0$ mJy.  This source is detected in all four \WISE\ bands (see Fig.~\ref{fig:postage2}), but its mid-IR SED is flatter than those of the others.  When combined with the photometry from the United States Naval Observatory B1.0 Catalog \citep{monet03}, the optical to mid-IR SED indicates the presence of a foreground galaxy at $z\sim 0.65$ (see Fig.~\ref{fig:SEDs}), likely lensing the background SMG.  The single line detected by the RSR at 92.826 GHz is identified as the redshift CO (2--1) line at $z=1.4820\pm0.0004$ when analysed together with its panchromatic SED, as shown in Fig~\ref{fig:photoz}.  Its relatively low redshift and modest brightness in the \Herschel\ bands indicates that PJ160722.6 is a comparatively low luminosity ($L_{IR}\approx 2\times10^{13}L_\odot$) object, similar to PJ142823.9 discussed above.  A faint NVSS 1.4 GHz radio source ($S_{1.4GHz}=1.3\pm0.5$ mJy) is also found on the \Herschel\ source position (see Fig.~\ref{fig:postage2}).

\subsection{PJ160917.8}
PJ160917 is another bright \Herschel\ SPIRE source in the ``Must-Do" survey and a ``350 \micron\ peaker" with $S_{350\mu} = 693$ mJy.  It has a bright \WISE\ counterpart with a relatively flat SED across its four bands, and it also has a relatively bright optical counterpart in SDSS J160918.25+604522.2 with $z_{ph}=0.43$ (see Fig.~\ref{fig:postage2}).   A bright AzTEC source with $S_{1100\mu}=73.8\pm7.0$ mJy, unresolved by the 8.5\arcsec beam, is detected with a high $S/N$, coincident with the \Herschel\ source position (see Fig.~\ref{fig:aztec}).  Our RSR spectrum shows 2 bright lines (see Fig.~\ref{fig:RSR}), unambiguously identifying it as the highest redshift object in our sample at $z=3.2557\pm0.0003$.  \citet{canameras15} report just one CO line at ``$z=3.3$", and their 2\arcsec resolution 870 \micron\ SMA image shows a classic lensing morphology of a partial arc, around the presumed lensing galaxy at  $z\approx0.5$.  The 1.4 GHz FIRST VLA source with $S_{1.4GHz}=1.50\pm0.30$ mJy coincides with the SMA source position.  

\subsection{Summary on Individual Sources}
The well-studied object with the most extensive complementary data, PJ020941.3, is clearly shown to be a high redshift hyperluminous IR galaxy lensed by a foreground galaxy.  Another well-studied object PJ142823.9 has the smallest apparent luminosity among our sample and has little evidence to support a lensing scenario.  Among the five \Planck-\Herschel\ sources with optical and SMA continuum imaging by \citet{canameras15}, PJ160917.8 is the clearest example of a strongly lensed object, while evidence for lensing is not as clear for the remaining four.  More sensitive, higher resolution imaging will be necessary to establish the lensing scenario for these remaining sources.

\section{Discussion}

\begin{figure}
\includegraphics[width=0.99\columnwidth]{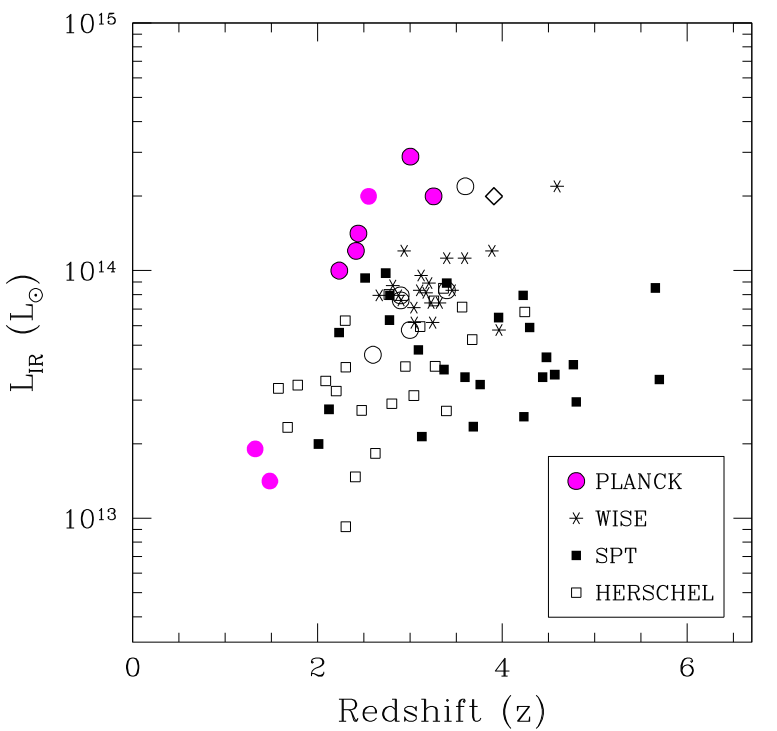}
\caption{Comparison of IR (8-1000 \micron) luminosity of the eight \Planck-\Herschel\ sources (filled circles) with those of strongly lensed SMGs identified by the SPT \citep{vieira13,weiss13} and the \Herschel\ \citep{bussmann13,wardlow13}.  Six additional \Planck\ sources reported by \citet{canameras15} are shown as empty circles.  The most luminous \WISE\ sources \citep{tsai15} are shown as stars, while the $z=3.91$ lensed infrared QSO APM 08279+5255 \citep{irwin98}, previously the most luminous object in the IR, is shown as a diamond.}
\label{fig:LFIR}
\end{figure}

\subsection{Extreme IR Luminosity of \Planck-\Herschel\ Sources}

\subsubsection{Sources with $L_{IR}\ge10^{14}L_\odot$}

The most remarkable aspect of these \Planck-\Herschel\ sources is that their IR (8-1000 \micron) luminosity is among the largest ever found.  As seen in the comparison of our IR luminosity with other luminous sources discovered by other recent IR surveys in Figure~\ref{fig:LFIR}, these  \Planck-\Herschel\ sources \citep[alongside the independently identified and overlapping targets by][]{canameras15} account for the majority of sources with rest-frame $L_{8-1000\mu}  \ge 10^{14} L_\odot$.   Nearly all of them either match or exceed the IR luminosity of the \WISE\ selected HyLIRGs by \citet{tsai15} that have recently been reported as ``the most IR luminous galaxies". Five out of the twenty \WISE\ sources reported by Tsai et al. have rest-frame $L_{IR} \ge10^{14} L_\odot$, with the $z=4.593$ source W2246$-$0526 being the most luminous with $L_{IR}=2.2\times10^{14} L_\odot$. In contrast, six out of eight \Planck-\Herschel\ sources in our sample and G145.2+50.9 in the Ca{\~n}ameras sample have $L_{IR}$ exceeding $10^{14} L_\odot$, and our brightest source, PJ105353.0 with $L_{IR}=2.9\times10^{14} L_\odot$, is the brightest IR galaxy ever discovered.  The strongly lensed $z=3.91$ QSO APM~08279+5255 \citep{irwin98,weiss07} with $L_{IR} = 2 \times 10^{14} L_\odot $ was the only known source with this type of IR luminosity previously, and these new \Planck\ and \WISE\ surveys have dramatically increased the number of objects with this type of IR luminosity, by more than an order of magnitude.  

Also shown in Figure~\ref{fig:LFIR} is the comparison of IR luminosity for our sample of \Planck-\Herschel\ sources with the strongly lensed SMGs discovered by \Herschel\ \citep{bussmann13,wardlow13} and the South Pole Telescope \citep[\SPT,][]{vieira13,weiss13}.  Their large IR luminosity ($L_{IR}=10^{12-14} L_\odot$) assisted by gravitational lensing made them popular targets of recent studies of luminous IR galaxy and dusty starburst phenomenon at $z\ge2$, during the epoch of galaxy build-up.  This comparison shows that {\em our \Planck-\Herschel\ sources are on average 5-10 times more luminous than these \Herschel\ and \SPT\ sources}, and they are the most luminous of the IR galaxies population that accounts for a large fraction of the cosmic star formation history \citep{lefloch05,caputi07,magnelli11}.  We note that two of our \Planck-\Herschel\ sources (PJ142823.9 \& PJ160722.6) as well as five of the Ca{\~n}ameras sample sources have IR luminosity comparable to these \Herschel\ and \SPT\ sources, demonstrating that the \Planck\ section is sensitive to these ``lower luminosity" ($L_{IR}\approx10^{13} L_\odot$) objects as well.  None of the \Planck\ selected sources so far are at $z>4$, unlike some of the \SPT\ or \Herschel\ sources, and this can be accounted by the small sample size and the comparatively shorter wavelength sample selection.


\begin{figure}
\includegraphics[width=0.99\columnwidth]{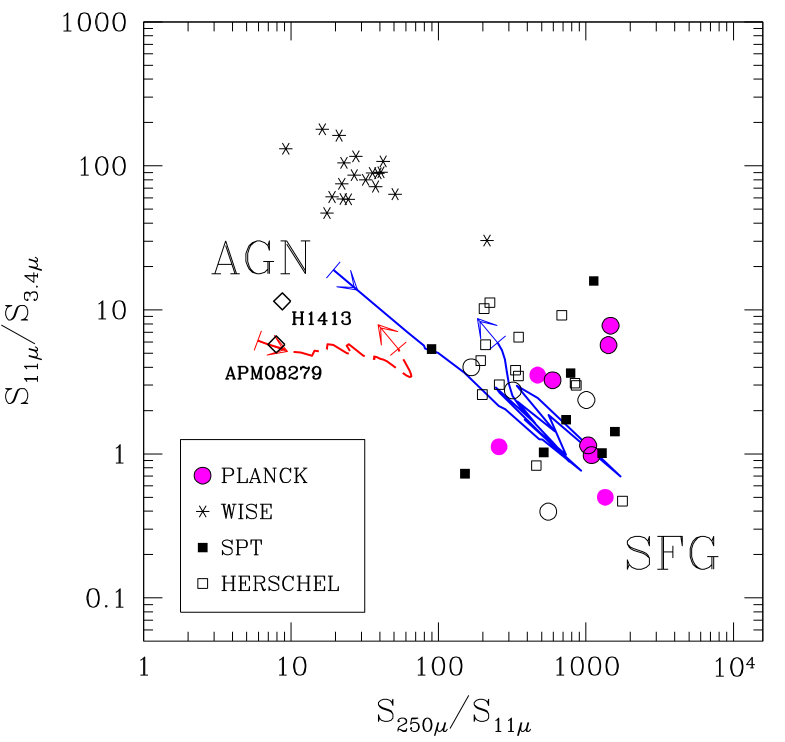}
\caption{Mid-IR to far-IR colour diagnostic plot for star forming and AGN galaxies by \citet{kirkpatrick13}, adopted for \WISE\ photometry instead of \Spitzer\ photometry.  All symbols are the same as Fig.~\ref{fig:LFIR}.  Another IR QSO H1413+117 is also shown using a diamond symbol in order to illustrate the parameter space occupied by AGNs.  Colour tracks for the  $z=2$ SF template (solid blue line) and the featureless power-law AGN template (long dashed red line) by Kirkpatrick et al. are shown, starting from $z=0.5$ and ending at $z=4$, as marked by the arrows.  The \WISE\ photometry for all sources shown are obtained from the NASA/IPAC Infrared Science Archive (IRSA).  The 3.4 \micron\ flux is that of the lensing galaxy in some cases (e.g., see Fig.~\ref{fig:SEDs}), leading to a large scatter in the $S_{11\mu}/S_{3.4\mu}$ ratio for the star forming galaxies.
}
\label{fig:colourcolour}
\end{figure}

\subsubsection{Nature of their high luminosity: AGN or star formation?}
\label{sec:SBAGN}

In the local Universe, the fraction of galaxies with a dominant AGN contribution increases with IR luminosity \citep[e.g., ][]{sanders96}, and the same trend may also hold true in the earlier epochs.  Many of the high redshift galaxies identified by early IR surveys, such as the Cloverleaf (H1413+117) and IRAS~F10214+4724 \citep{barvainis92,barvainis95}, are indeed hosting a luminous, dust-obscured AGN.  Therefore, it is natural to suspect ubiquitous AGN activity among these \Planck-\Herschel\ sources, especially given their extremely large IR luminosity ($L_{IR} \ge10^{14} L_\odot$) typically associated with QSOs.  Detailed follow-up studies of the \SPT\ and \Herschel\ sources with $L_{IR} \ge10^{13} L_\odot$ have concluded that many are gravitationally lensed sources with a typical magnification factor of $\mu\approx 10$  \citep{bussmann13,vieira13,dye14}.  If some or all of the \Planck-\Herschel\ sources are also magnified by a similar factor \citep[$\mu\approx10$ for PJ020941.3, ][]{geach15}, then their intrinsic IR luminosity is $L_{IR} \ge10^{13} L_\odot$, still in the QSO regime.

A quantitative analysis of the AGN contribution to the total IR luminosity is possible if the mid- to far-IR SED is fully mapped.  By analyzing a large sample of high redshift galaxies with extensive mid- and far-IR data from \Spitzer\ and \Herschel, \citet{kirkpatrick12,kirkpatrick13} have shown that the energetic contribution by AGN activity can be determined quantitatively through a spectral decomposition, even for heavily obscured, Compton thick sources.  In particular, \citet{kirkpatrick13} have shown that combining the longer wavelength photometry from \Herschel\ with mid-IR photometry from \Spitzer\ (or \WISE) yields the most reliable separation of AGN and SF dominated galaxies since they probe the widest range of dust properties affected.  

A modified version of this mid-IR to far-IR colour diagnostic plot, replacing \Spitzer\ photometry bands with \WISE\ photometry bands, is shown in Figure~\ref{fig:colourcolour}, and our \Planck-\Herschel\ sources appear in the region dominated by star formation activity.  The two IR QSOs with the characteristic broad emission lines, APM~08279+5255 \citep{irwin98} and the Cloverleaf \citep[H1413+117,][]{barvainis92,barvainis95} are located near the colour track of the featureless power-law AGN (long dashed line), as expected.  All but one of the luminous \WISE-selected IR galaxies have a relatively flat mid- to far-IR flux ratio ($S_{250\mu}/S_{11\mu}$) characteristic of power-law AGNs, but with systematically redder mid-IR colour ($S_{11\mu}/S_{3.4\mu}>30$).  By analysing their SEDs, \citet{tsai15} have concluded that these \WISE-selected sources are likely powered by highly obscured AGNs, with their hot dust ($T_d\sim 450$ K) accounting for the high luminosities.  Their mid-IR colour is systematically redder than the power-law AGN model track, but this is a consequence of the ``W1W2-dropout" selection in their sample definition. In contrast, nearly all of our \Planck-\Herschel\ sources \citep[also by ][]{canameras15} as well as all of the \SPT- and \Herschel-selected IR sources appear along the SF template track (solid line). \citet{kirkpatrick15} have shown that AGN-powered sources rapidly cross over from the SF dominated to the AGN region on this diagnostic plot once the AGN contribution exceeds 60\% of the total luminosity, and {\em this analysis clearly indicates that our \Planck-\Herschel\ sources are primarily powered by SF activity}.  This conclusion is further supported by the fact that nearly all \Planck-\Herschel\ sources also follow the radio-IR correlation characteristic of star forming galaxies (see \S~\ref{sec:SBSED} and Fig.~\ref{fig:SEDs}), and the nature of the energetic activities powering their large luminosity is substantially different between the \Planck-selected and the \WISE-selected SMGs, despite their similarly extremely IR luminosity.

\subsection{Molecular Gas Mass Estimates}

\subsubsection{Gas Mass Estimates from CO Line Luminosity \label{sec:COmass}}

A standard method for deriving a total molecular gas mass from a redshifted CO line measurement is to adopt a line ratio between different rotational transitions to estimate the CO (1--0) transition luminosity $L'_{CO(1-0)}$ and then to translate this line luminosity to a total molecular gas mass assuming an ``$\alpha_{CO}$" conversion factor \citep[see the review by][]{carilli13}.  The Table~2 by \citet{carilli13} gives the average CO line ratios for SMGs as $L'_{CO(2-1)}/L'_{CO(1-0)}$, $L'_{CO(3-2)}/L'_{CO(1-0)}$, and $L'_{CO(4-3)}/L'_{CO(1-0)}$ of 0.85, 0.66, and 0.46, respectively.  Using these ratios, we can convert the measured CO line luminosities reported in Table~\ref{tab:RSR}  to $L'_{CO(1-0)}$ and then to  $M_{H2}$ using the ``Galactic" conversion factor $\alpha_{CO} \equiv M_{H2}/L'_{CO(1-0)}=4.3 M_\odot \, [{\rm K\, km s^{-1}\, pc^{-2}}]^{-1}$ \citep{bolatto13}.

The derived total molecular gas masses of our eight \Planck-\Herschel\ sources, uncorrected for the unknown magnification factor $\mu$,  range between $\mu M_{H2} = 5.4\times 10^{11}M_\odot$ and $7.8\times 10^{12}M_\odot$.  For a typical magnification factor of $\mu\sim10$, the intrinsic molecular gas masses range between $(0.5-8)\times 10^{11} M_\odot$, which are near to top of the range of gas masses found for high redshift SMGs that are not strongly lensed \citep[see][]{carilli13}.  The measured line widths (FWHM) of 400-650 \kms\ are somewhat large, but they are fairly typical for the most gas-rich SMGs (see Fig.~5 by Carilli \& Walter).  We note that earlier studies often used the ``ULIRG" conversion factor of $0.8 M_\odot [{\rm K\, km s^{-1}\, pc^{-2}}]^{-1}$, leading to systematically smaller total gas mass, without a clear physically motivated justification.  Following \citet{scoville15}, who have argued for using the Galactic conversion even for high redshift SMGs after a thorough review of the topic, we also adopt the Galactic conversion factor for our molecular gas mass estimates.  We revisit this decision by comparing these total molecular gas mass estimates with the dust continuum based ISM mass estimates in the next section (\S~\ref{sec:ISMmass}).


An upper bound to the total mass requires that the total gas mass cannot exceed the dynamical mass of the system.  A virial mass of a pressure supported spherical system with a 1-D velocity dispersion ($\sigma\equiv FWHM/8ln2)$ is $6\times10^9[\frac{FWHM}{400\,{\rm km\, s}^{-1}}]^2 R_{kpc} M_\odot$, where $R_{kpc}$ is the characteristic radius of the system in kpc.  For a rotationally supported disk with a radius $R_{kpc}$, an equivalent dynamical mass with a rotational speed $V_{circ}\equiv FHWM/2$ is $9\times10^9[\frac{FWHM}{400\,{\rm km\, s}^{-1}}]^2 R_{kpc} M_\odot$.  Therefore, the maximum possible gas mass for a 10 kpc radius disk is $2.4\times 10^{11} M_\odot$ for a galaxy with observed CO line width of 650 \kms, which exceeds the CO-derived gas mass for the eight \Planck-\Herschel\ sources by factors of 2 to 30.  These gas mass estimates may be somewhat over-estimated, by a factor of order unity as discussed below.  Accounting for the remainder of the difference requires that nearly all of these galaxies {\em must be} strongly lensed by at least these factors (2 to 30), even though we lack any direct evidence for lensing in most cases.  The empirical estimate for the magnification factor based on measured CO line width and line luminosity, as proposed by \citet{harris12}, is essentially a different formulation of this upper mass limit.

\begin{figure}
\includegraphics[width=0.9\columnwidth]{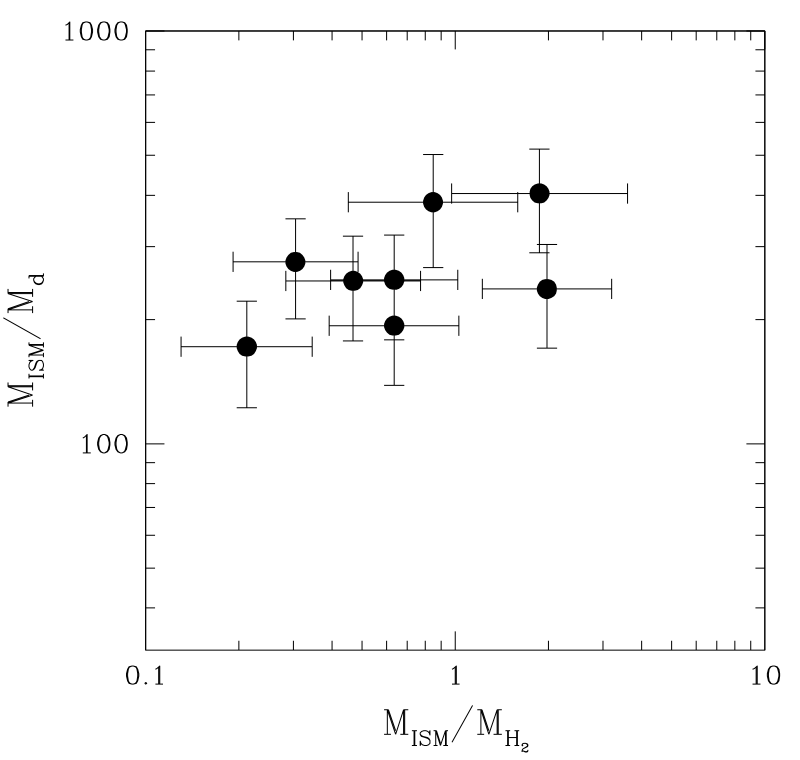}
\caption{A comparison of gas-to-dust ratio (GDR; $M_{ISM}/M_d$) and the ratio of gas masses derived from AzTEC 1100 \micron\ flux ($M_{ISM}$) to the CO-derived $M_{H2}$.  The median GDR derived for the eight \Planck-\Herschel\ sources is $\approx 250$, while the median $M_{ISM}/M_{H2}$ ratio is $\approx 0.65$.}
\label{fig:compareM}
\end{figure}

\subsubsection{ISM Mass Estimates from AzTEC $1100\,\mu$m Photometry \label{sec:ISMmass}}

As an alternative to deriving total gas mass from CO measurements that can be observationally expensive, \citet{scoville15} have proposed an empirical calibration for total ISM mass derived from the optically thin, long wavelength dust continuum emission and adopting a gas-to-dust ratio (GDR).  As shown in their Figure~1, CO line luminosity $L'_{CO}$ is linearly correlated with the monochromatic luminosity $L_{850\mu}$ at the rest frame 850 \micron\ for local star forming galaxies, local ULIRGs, and $z\sim2$ SMGs, spanning over six orders of magnitude in luminosity.  This in turn can be used to derive a dust mass using the constant scaling factor $\alpha_{850\mu} = L_{850\mu} / M_{d}$ based on the local calibration.

Our AzTEC 1100 \micron\ photometry is well suited for this application as we are probing near the Rayleigh-Jeans part of the spectrum ($\nu_{rest}$ = 260-470 \micron) at the redshift of these \Planck-\Herschel\ sources (see Fig.~\ref{fig:SEDs}).  Considering that the average IR luminosity density should be at least 10 times higher for our \Planck-\Herschel\ sources compared with the local and $z\sim2$ galaxies used for the calibration, we adopt a dust temperature 1.5 times higher than considered by Scoville et al., or $T_d=35$ K\footnote{This adopted dust temperature is lower than the the luminosity-weighted dust temperature derived from the IR SED analysis ($T_d$ = 42-84 K, see Table~\ref{tab:chisq}), but \citet{scoville15} make a compelling argument for using a lower, mass-weighted $T_d$ -- see their Appendix A.2 for a detailed discussion.}.
Adopting Eq.~A14 by Scoville et al., the ISM mass, which is effectively the total molecular gas mass, can be derived from the observed AzTEC flux density as,
	$$M_{ISM}(M_\odot) = \frac{4.74\times 10^4  D_{L}^{2}}{(1+z)^{4.8}}[\frac{S_{1100\mu}}{\rm mJy}]  [\frac{6.7\times10^{19}}{\alpha_{850\mu}}] [\frac{\Gamma_{RJ}}{\Gamma_{0}}],$$
	where $D_L$ is the luminosity distance in Mpc, and $S_{1100\mu}$ is the measured flux density at 1100 \micron\ in mJy (for $\lambda_{rest} \ge 250$ \micron).  Here we adopt the value of $\alpha_{850} = 6.7 \times 10^{19}$ ergs/sec/Hz/ M$_\odot$ (Scoville et al. Eq.~A11), and the RJ correction factor $\Gamma_{RJ}$ that accounts for any deviation away from the $\nu^{2}$ dependance as 
  $$\Gamma_{RJ}(T_{d},\nu_{obs}, z) = \frac{h\nu_{obs} (1+z)/kT_{d}}{exp({h\nu_{obs}(1+z) /kT_{d}}) -1}.$$

The total ISM mass we derive for these \Planck-\Herschel\ sources range between $\mu M_{ISM}=5.0\times 10^{11} M_\odot$ and $5.1\times 10^{12}  M_\odot$, uncorrected for the unknown magnification factor $\mu$ (see Table~\ref{tab:chisq}).   As shown in Figure~\ref{fig:compareM}, the median ratio between this ISM mass and the CO-derived $M_{H2}$ mass (\S~\ref{sec:COmass}) for our sample is $\left< M_{ISM}/M_{H2}\right> \approx 0.65$, with a significant scatter, and this indicates that the ISM mass derived from the AzTEC 1100 \micron\ photometry is systematically smaller by a factor of 1.5 compared with the CO-derived $M_{H2}$.  The sample plot also shows that the gas-to-dust ratio for the same sample has a median value of $\left< M_{ISM}/M_d\right> \approx 250$, which is close to the nominal value in the local Universe for metal-rich galaxies \citep[$\approx170$,][]{draine07}.  This median value and the small scatter suggest that the ISM masses derived following calibration by \citet{scoville15} and the dust masses derived using an independent calibration (see Appendix~\ref{sec:Md}) are at least consistent, probably because of their shared assumptions. 

A possible explanation that our CO-derived $M_{H2}$ might be a slight over-estimate is in line with the results of recent theoretical works.  By modeling the impact of metallicity, gas temperature, and velocity dispersion, \citet{narayanan12} have postulated a weakly dependent reduction in $\alpha_{CO}$ with increasing velocity dispersion for high redshift galaxies with a larger turbulent support.  Specifically, Narayanan et al. model predicts a reduction in $\alpha_{CO}$ by 20\% and 40\% (compared with the Galactic value) with a factor of 2 and 5 increase in velocity dispersion, respectively.  A reduction in $\alpha_{CO}$ by this magnitude would bring our $M_{ISM}/M_{H2}$  ratio and the gas-to-dust ratio to a much better agreement.  We note that our estimates of $M_{H2}$ are systematically larger than those reported by \citet{canameras15} by a significant amount, with the median for the five objects in common being 6.6 times larger for our estimates.  In their gas mass derivation, Ca{\~n}ameras et al. adopted the ``ULIRG" conversion factor, which immediately leads to 5 times smaller gas masses.  In addition, Ca{\~n}ameras et al. did not explicitly account for the excitation dependent correction for observing different CO rotational transitions \citep[see][]{carilli13}, leading to a further factor of 1.5 reduction in their $H_2$  mass estimates.  Their combined effects should lead to 7.5 times smaller $H_2$ masses, which can more than account for the factor of 6.6 difference.  The lower gas masses derived by Ca{\~n}ameras et al. should in turn lead to an average gas-to-dust ratio of $GDR\approx35$\footnote{\citet{canameras15} report gas-to-dust ratios of $GDR = 32-112$ using their CO-derived $M_{H2}$ and dust mass derived from their SED modeling.  This low $GDR$ suggests that either they under-estimate their gas masses or over-estimate the dust masses.  In deriving their dust masses, Ca{\~n}ameras et al. use the {\em luminosity-weighted} $T_d$, which should lead to systematically lower estimates of $M_d$, as explicitly cautioned to avoid  by \citet{scoville15}.   Therefore, it follows that their $M_{H2}$ estimates are indeed substantially under-estimated.}.  These considerations of average GDR and $M_{ISM}/M_{H2}$ ratio suggest that our CO-derived gas masses are within a factor of 2 of the correct values.

\section{Conclusions}

A further understanding of how to link the IR galaxies making up the cosmic infrared background (CIRB) with the entire galaxy population across all redshifts is revolutionized by the systematic detection of gravitationally lensed SMGs. 
These systems are some of the most interesting objects to study because the magnifying property of lensing allow us to probe (1) physical details of the intense star formation activities at sub-kpc scale that are beyond the conventional means including the ALMA and the JVLA; and (2) dusty star forming galaxies beyond the current detection limit, with $SFR=10-100\, M_\odot$ yr$^{-1}$ that account for the bulk of stellar mass build-up at $1<z<4$.  

In this pilot study for identifying a large population of strongly lensed SMGs using the \Planck\ survey data, we report our successful identification of eight candidate sources drawn from the \Planck\ Compact Source Catalog using the public \Herschel\ archival data and the results from our detailed follow-up investigation utilizing new AzTEC 8.5\arcsec resolution 1100 \micron\ imaging  and CO spectroscopy using the Redshift Search Receiver on the LMT.  Our new observations have shown that these are some of the most luminous galaxies ever found with $L_{IR}$ up to $3\times 10^{14} \mu^{-1} L_\odot$, and our \Planck-\Herschel\ sources at redshifts of $1.3<z_{CO}<3.3$ are also some of the most gas-rich galaxies found ($M_{H2} = (0.6-7.8)\times 10^{11} M_\odot$ for $\mu\approx10$).  The SED analysis using the mid- to far-IR colour diagnostic plot developed by \citet{kirkpatrick13} shows that these sources are powered primarily by the intense SF activity, unlike the similarly luminous IR galaxies identified in the \WISE\ survey by \citet{tsai15}, which are powered primarily by luminous AGNs.  Our examination of the radio-IR correlation also supports this SF scenario.

The results of our pilot study suggest that a much larger population of strongly lensed SMGs can be potentially found in the \Planck\ Compact Source Catalog, including sources with $L_{IR}$ down to $\sim10^{13} \mu^{-1} L_\odot$.  At least four (out of eight) \Planck-\Herschel\ sources are associated with a bright foreground optical galaxy, strongly supporting the lensing hypothesis, and the $z=2.554$ source PJ020941.3 has been definitively shown to be a partial Einstein ring lensed by a $z=0.2$ foreground galaxy \citep[$\mu\approx13$,][]{geach15}.  A larger survey to identify a much larger sample of lensed SMGs in the  \Planck\ Compact Source Catalog is already underway, and the results of this expanded study will be reported in a future paper.

\section*{Acknowledgments}
We would like to thank A. Pope, N. Scoville, R. Snell, J. Vieira for useful discussions and D. F. Gallup for his assistance during the initial development of the project.  
This work would not have been possible without the long-term financial support from the Mexican Science and Technology Funding Agency, CONACYT (Consejo Nacional de Ciencia y Tecnolog\'{i}a) during the construction and early operational phase of the Large Millimeter Telescope Alfonso Serrano, as well as support from the the US National Science Foundation via the University Radio Observatory program, the Instituto Nacional de Astrof\'{i}sica, \'{O}ptica y Electr\'{o}nica (INAOE) and the University of Massachusetts, Amherst (UMass). The UMass LMT group acknowledges support from NSF URO and ATI grants  (AST-0096854, AST-0215916, AST-0540852, and AST-0704966) for the LMT project and the construction of the RSR and AzTEC.  KH and RC would like
to acknowledge support from a William Bannick Student Travel Grant.  We are grateful to all of the LMT observers from Mexico and UMass who took data for this project. This publication makes use of data products from the Wide-field Infrared Survey Explorer, which is a joint project of the University of California, Los Angeles, and the Jet Propulsion Laboratory/California Institute of Technology, funded by the National Aeronautics and Space Administration. This work is based in part on observations made with the Herschel Space Observatory, which is an ESA space observatory with science instruments provided by European-led Principal Investigator consortia and with important participation from NASA, and the Planck, which is European Space Agency mission with significant NASA involvement.  This research has made use of the NASA/ IPAC Extra- galactic Database (NED) which is operated by the Jet Propulsion Laboratory, California Institute of Technology, under contract with the National Aeronautics and Space Administration. This research has made use of the NASA/ IPAC Infrared Science Archive, which is operated by the Jet Propulsion Laboratory, California Institute of Technology, under contract with the National Aeronautics and Space Administration.

\bibliography{references}

\newpage

\appendix

\section{Determining Unique CO Redshift using IR Photometric Redshift}
\label{sec:appendixRSR}

\begin{figure*}
\includegraphics[width=5.cm]{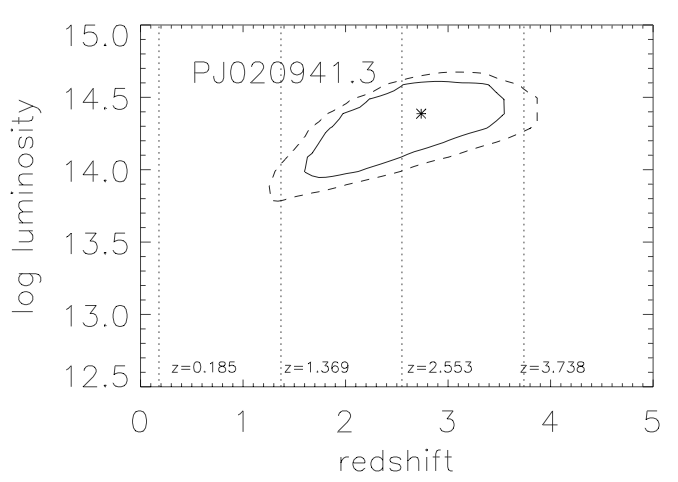}
\includegraphics[width=5.cm]{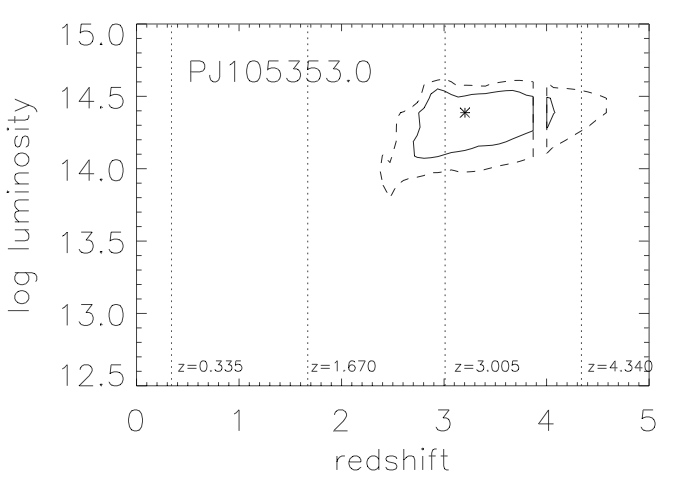}
\includegraphics[width=5.cm]{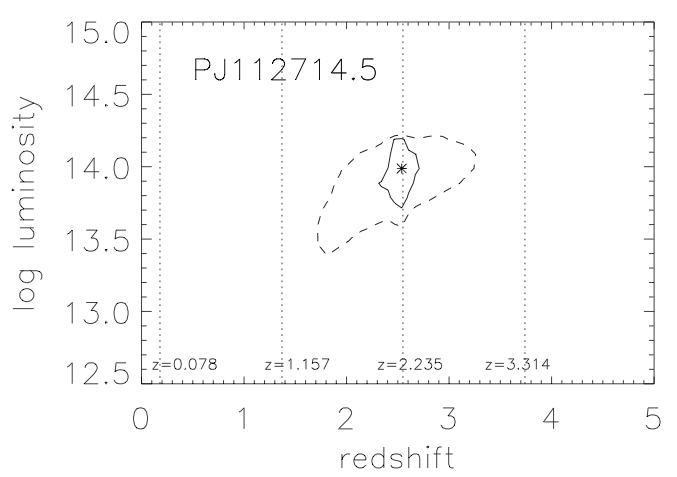}
\includegraphics[width=5.cm]{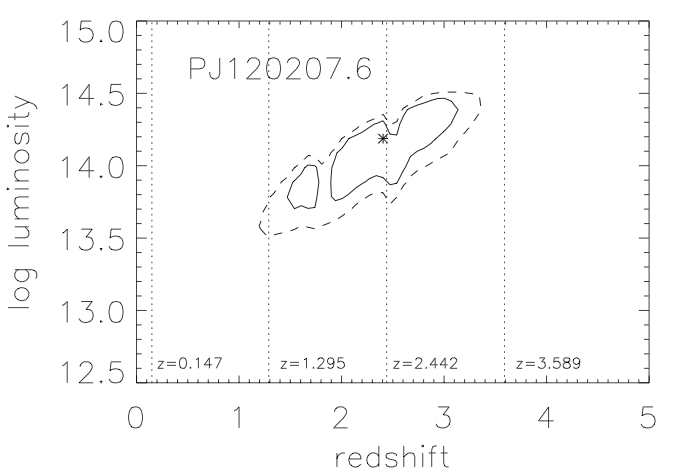}
\includegraphics[width=5.cm]{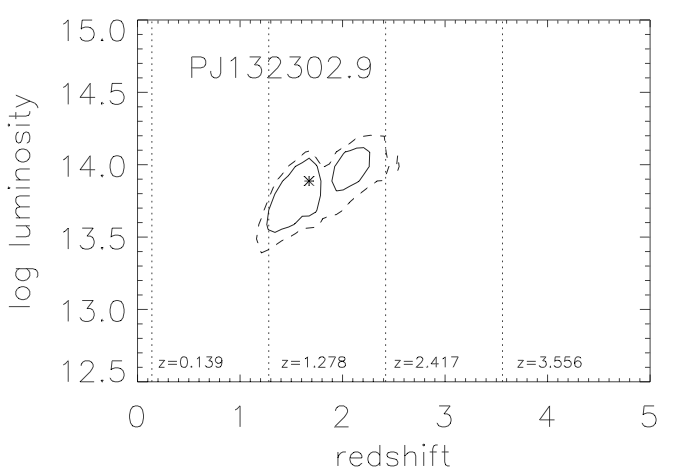}
\includegraphics[width=5.cm]{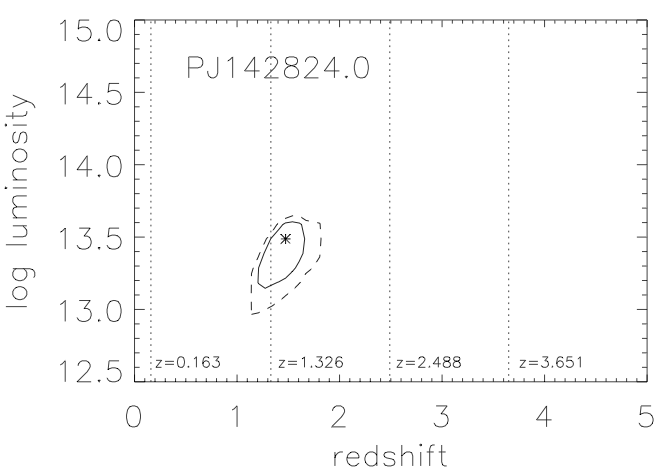}
\includegraphics[width=5.cm]{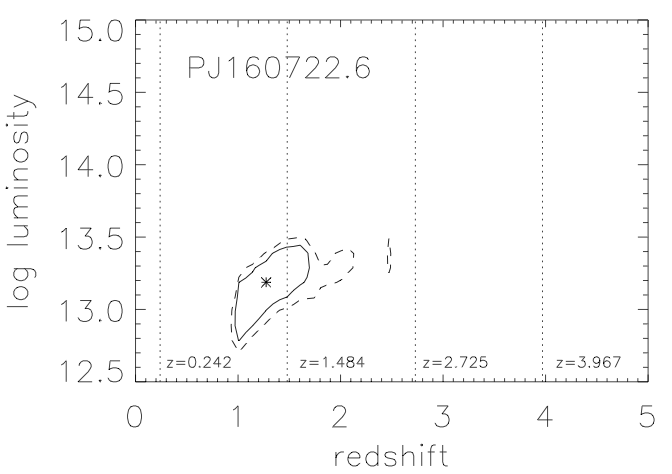}
\includegraphics[width=5.cm]{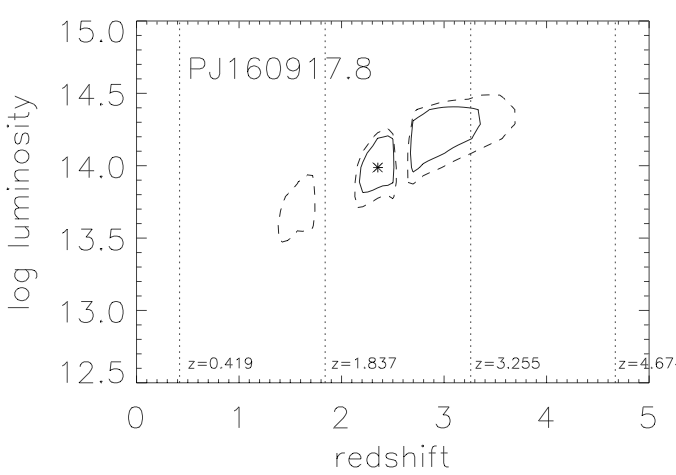}
\caption{A $\chi^2$ based panchromatic photometric redshift support for determining a unique CO redshift.  The minimum $\chi^2$ solution is marked by an asterisk, and the dotted and solid contour denote the 68 and 90 percentile confidence levels \citep{wall96}.  Discrete possible redshift solutions corresponding to the observed CO lines are shown as regularly spaced vertical lines in each panel.}
\label{fig:photoz}
\end{figure*}

A unique redshift can be determined when two or more spectral lines are detected in a single RSR spectrum, such as for PJ160917.8 in Fig.~\ref{fig:RSR}.  Only one CO line falls within RSR band in the redshift range between $z=0$ and $z\approx3.15$, however, and additional information such as photometric redshift is needed in order to distinguish different redshift solutions  -- see discussions by \citet{yun15}.  Specifically, Yun et al. have demonstrated the use of radio-millimetric photometric redshift technique \citep{carilli99} as an effective means to provide the redshift support using minimum additional data.  A more thorough photometric redshift analysis using all available photometry data can in principle provide a stronger support for the redshift discrimination.  

The use of a full photometric redshift analysis for deriving a unique redshift solution for an RSR spectrum is demonstrated in Fig.~\ref{fig:photoz}.  For this analysis, photometric data points from \WISE\  (11 \& 22 \micron), \Herschel, AzTEC, and VLA are analyzed using the template SEDs by \citet{efstathiou00}.  A minimum $\chi^2$ value for an entire suite of template SEDs is computed at each grid point in the redshift-luminosity parameter space, and a map of the minimum $\chi^2$  values is constructed.  The over-all best fit solution with the overall minimum $\chi^2$ is shown as an asterisk in each panel, and it is often close to one of the redshift solutions corresponding to a CO line seen in the RSR spectrum.   The dotted and solid contours denote the 68 and 90 percentile confidence levels derived based on the reduced $\chi^2$ values \citep{wall96}, and a weighted likelihood is used to select a unique redshift solution based on this analysis.  Vertical gaps and discontinuities in the confidence contours seen in some panels are the artifacts arising from the \WISE\ photometry points passing through the PAH emission line features in the SED templates, which are generally not modeled very well by the templates used.  These \WISE\ photometry provide useful constraint on the photometric redshift for high redshift dusty starbursts with their characteristic red colour \citep[see][]{yun08,yun12}, and including these data  is still helpful for narrowing down the acceptable solutions, even with these apparent artifacts.

\section{Line Width Correction for the Instrumental Resolution}
\label{sec:appendixLW}

Unlike nearby spiral galaxies that are rotationally supported, many high redshift CO sources are thought to be at least partially pressure-supported, and their CO line profiles are often characterized as a Gaussian in shape.  The frequency response of the RSR is also roughly Gaussian.  For a line with an intrinsic Gaussian width $\sigma_0$, the full width half maximum (FWHM) CO line width is $W_{0}=\sigma_0\sqrt{8ln2}$.  The measured CO line then has a Gaussian width of $\sigma_c$ with an apparent line width of $W_c=\sigma_c\sqrt{8ln2}$, which is the result of the intrinsic line width $W_0$ convolved with the Instrumental resolution of $R=\sigma_R\sqrt{8ln2}$.  Since $\sigma_c = \sqrt{\sigma_0^2 + \sigma_R^2}$, the intrinsic line width $W_0$ can be recovered as
\bea
W_0 & = & 8(ln2) \sigma_0 \\
        & = & 8(ln2) \sqrt{\sigma_c^2 - \sigma_R^2} \\
        & = & 8(ln2)\sigma_c \sqrt{1-(\frac{\sigma_R}{\sigma_c})^2} \\
        & = & W_c \sqrt{1-(\frac{R}{W_c})^2}.
\eea
The spectral channel width of 31.25 MHz translates to a velocity resolution of $R=100\, (\frac{93.75}{\nu_{CO}})$ km s$^{-1}$, where $\nu_{CO}$ is the observed CO line frequency in GHz.  The CO line width $\Delta V$ reported in Table~\ref{tab:RSR} is the deconvolved line width $W_0$ after correcting for the instrumental spectral resolution of the RSR as shown here.  The magnitude of correction is small, $\le 30$ \kms, and is expected to be significantly only when the line width is close to the spectral resolution.

\section{Modified Blackbody Model}
\label{sec:appendixMBM}

	The dust mass estimates are highly dependent on the dust temperature distribution amongst the ISM. 	\citet{yun02} provide an expression for a radio-to-FIR spectrum parameterised by SFR, dust temperature, and luminosity distance in their Eq~15.  In order to characterize the dust temperature for each source we adapted this equation to its relative functional form with dust temperature as a free parameter. In modeling the Rayleigh-Jeans part of the dust spectrum, one can adopt a value for $\nu_{c} $, the critical frequency at which the light coming from the dust clouds becomes optically thick. We take $ \nu_{c} $ to be 2000 GHz (150 \micron) as in Yun \& Carilli. The dust emissivity index, $ \beta$, is held fixed at a value of 1.8 to remain consistent with the comparative analyses by \citet{scoville15}. Typical values for $ \beta$ with respect to SMGs range from 1.5 to 2.0 \citep[see][]{scoville14}. There are essentially two descriptors of the SED model that uniquely characterize the best fit once a given value for $\beta$ is assumed: a normalization factor related to the total star formation rate $SFR$, and the dust temperature $T_d$. 
	$$S_{d}(\nu) = \frac{(1.3 \times10^{-6}) (SFR)(1+z) \nu_0^{3}}{D_{L}^{2} [e^{0.048\nu_0/T_{d}} -1]}[1-e^{-( \nu_0/ \nu_{c})^{\beta}} ]$$
where $\nu_0\equiv(1+z)\nu$ is the frequency in the rest frame of the object.	A discrete range of normalization factors for each source were recorded for every dust temperature sampled (ranging from 0.5-99.5 K with increments of 0.4 K) in order to compute minimum reduced $\chi^2$ values. In order to track the internal degeneracy between the normalization factor and dust temperature and to effectively characterize the upper and lower limits, we produce a 2-dimensional plot given the normalization factor versus the rest-frame dust temperature.  

\begin{figure*}
\includegraphics[width=5.cm]{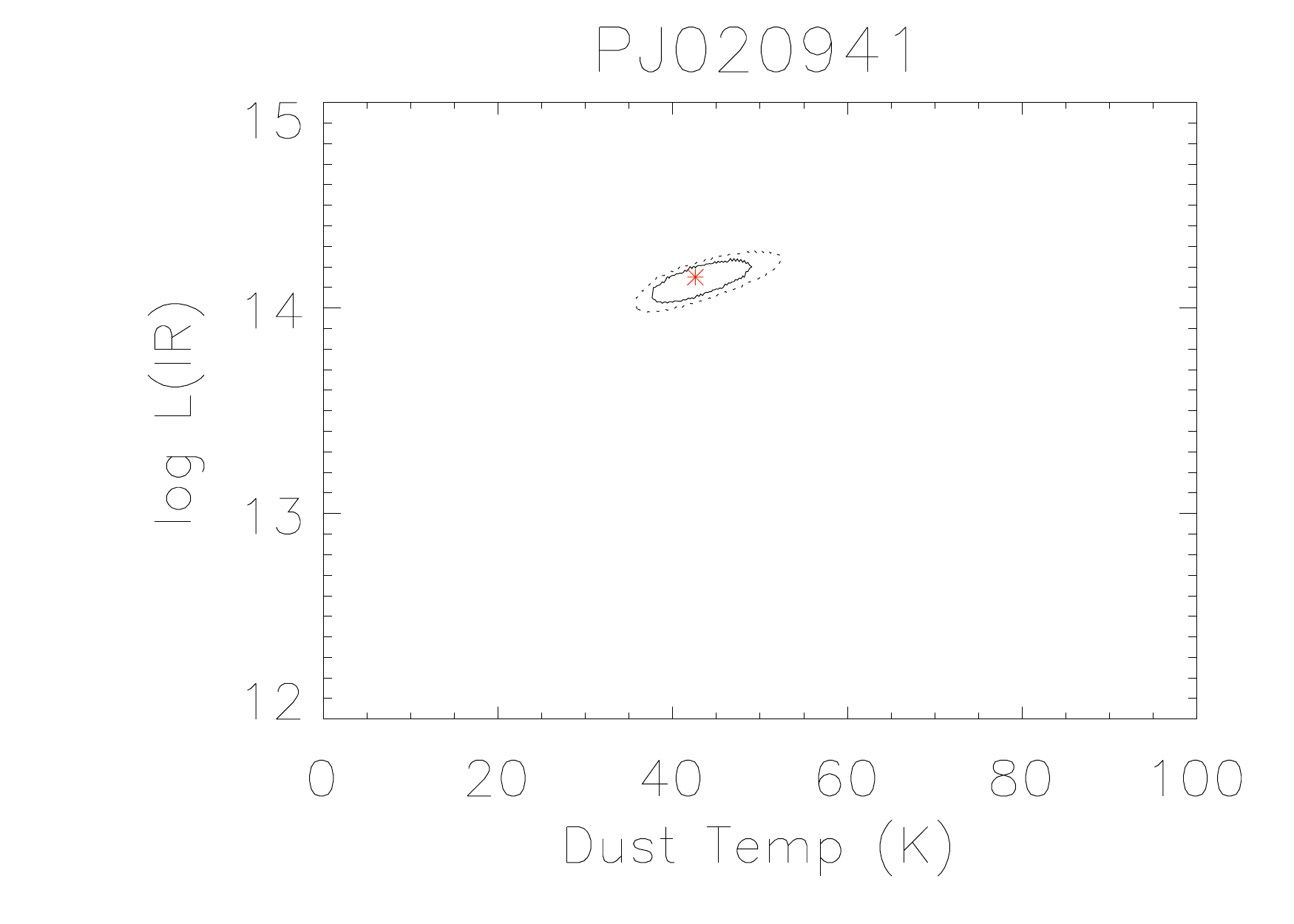}
\includegraphics[width=5.cm]{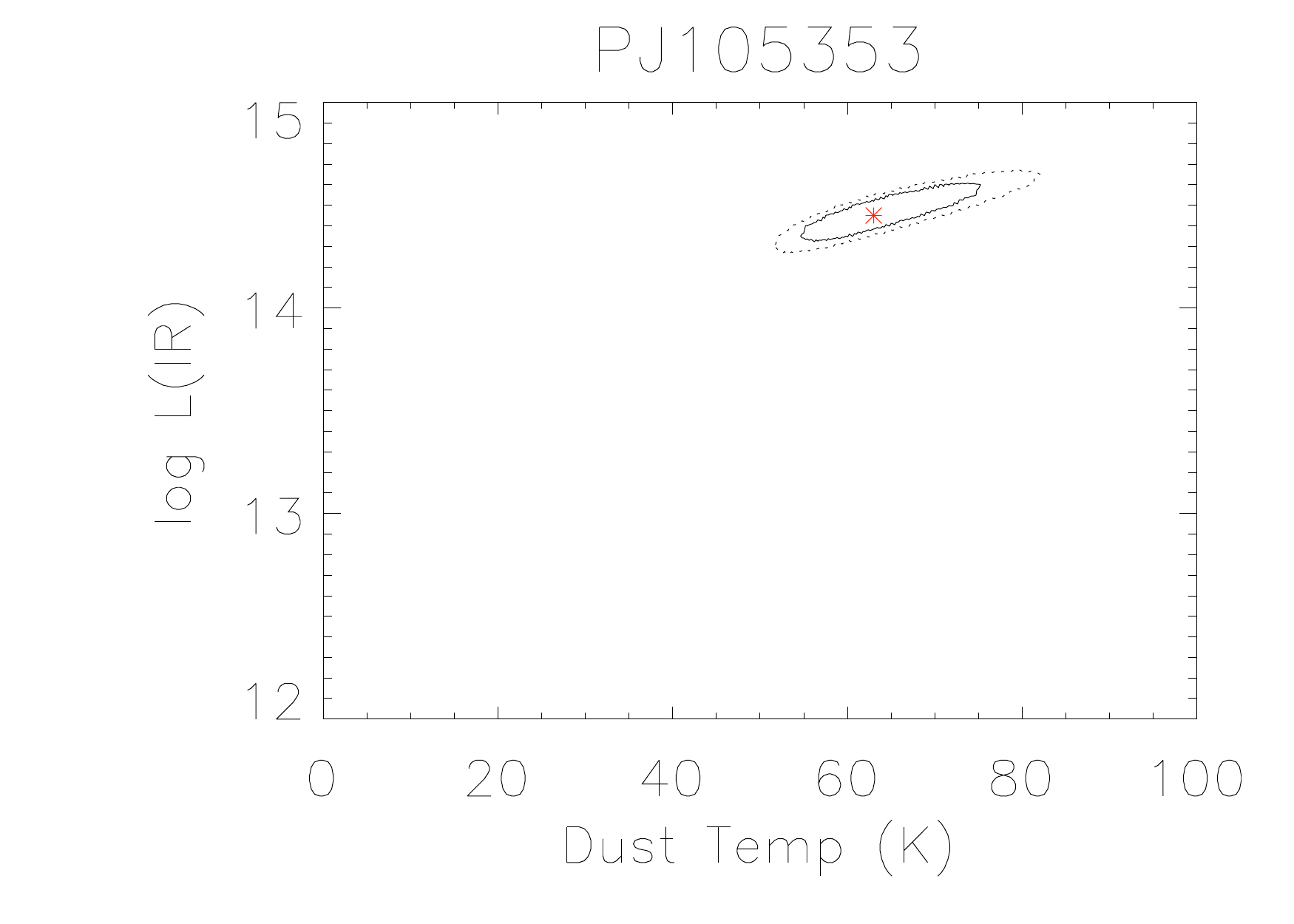}
\includegraphics[width=5.cm]{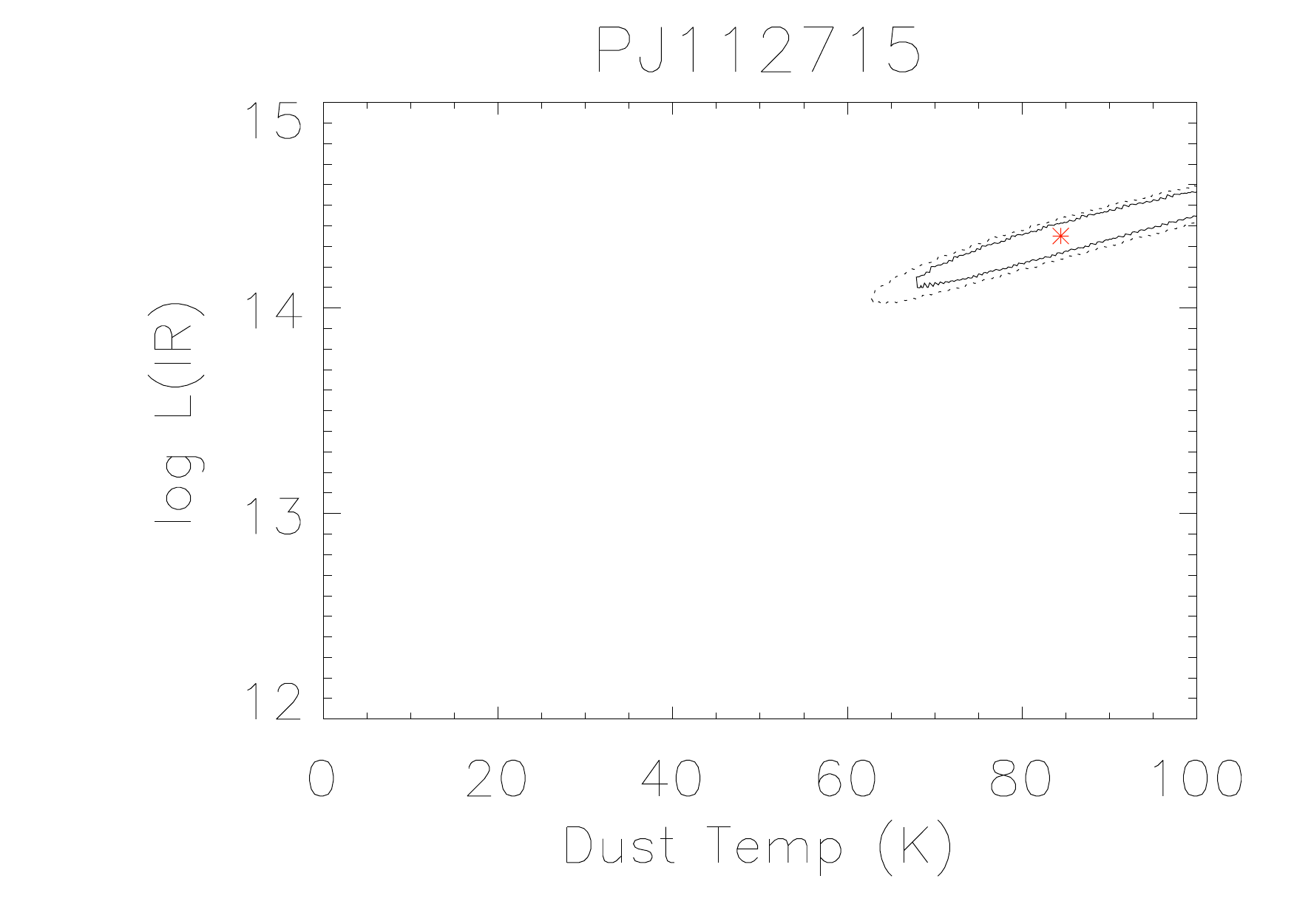}
\includegraphics[width=5.cm]{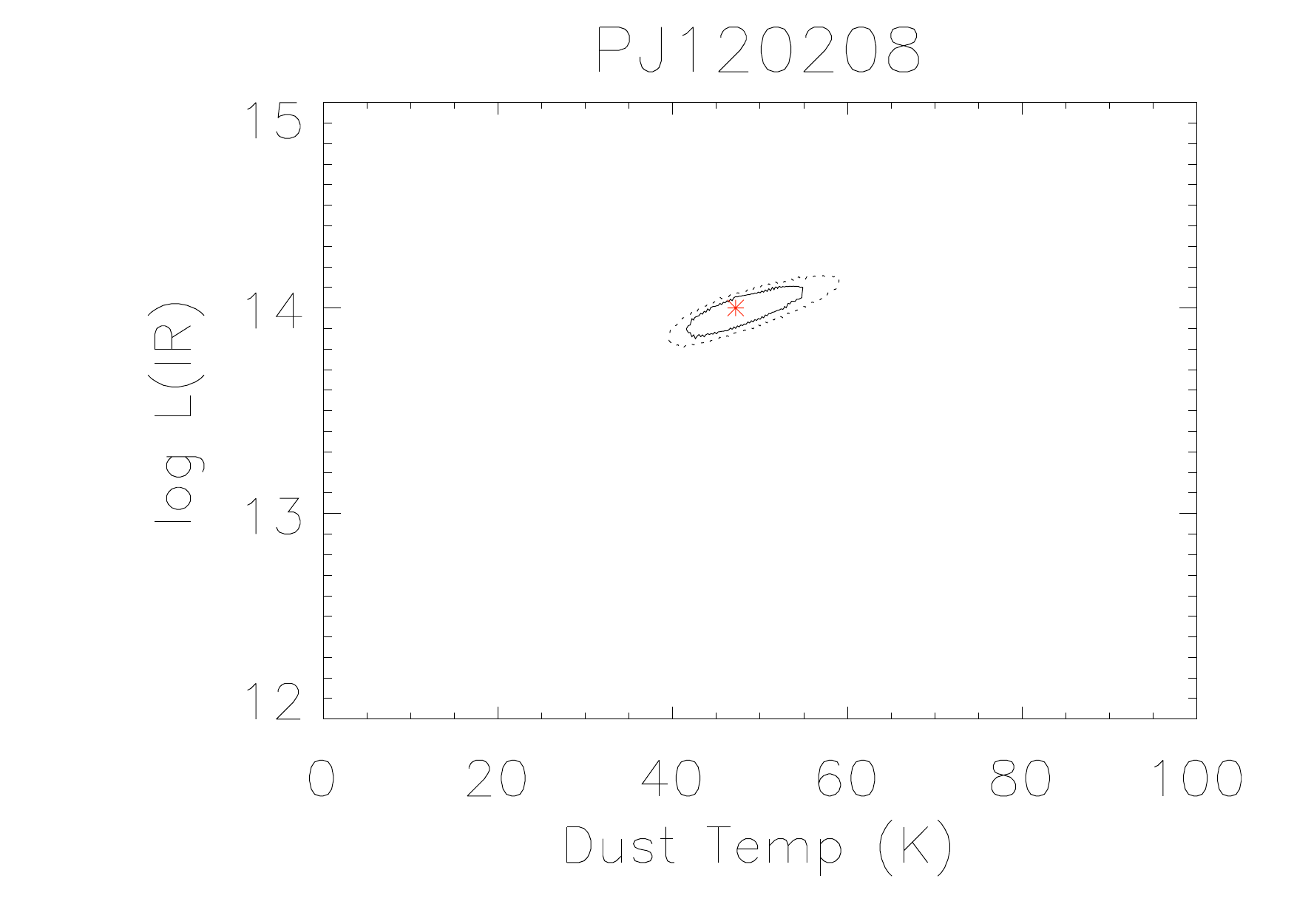}
\includegraphics[width=5.cm]{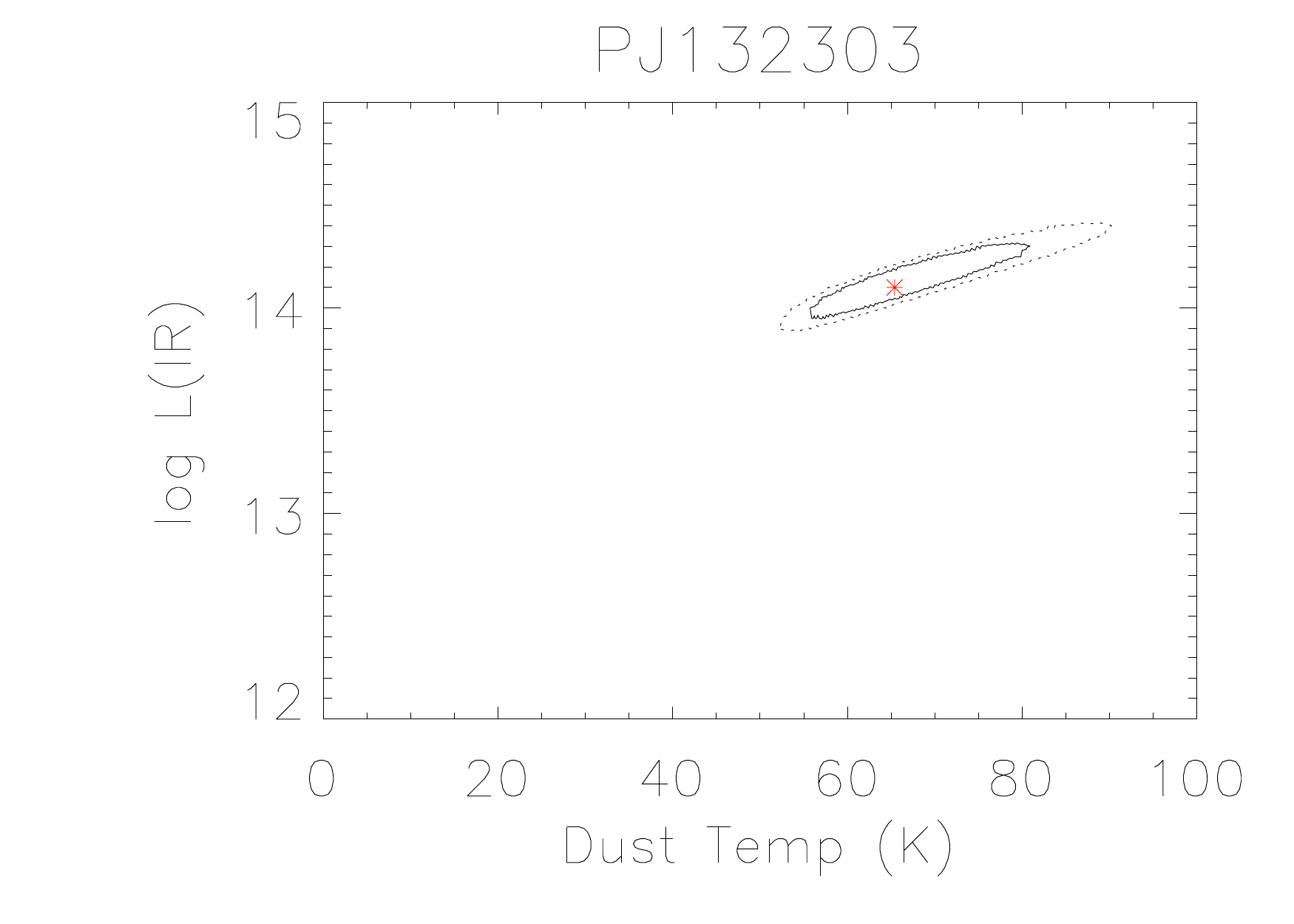}
\includegraphics[width=5.cm]{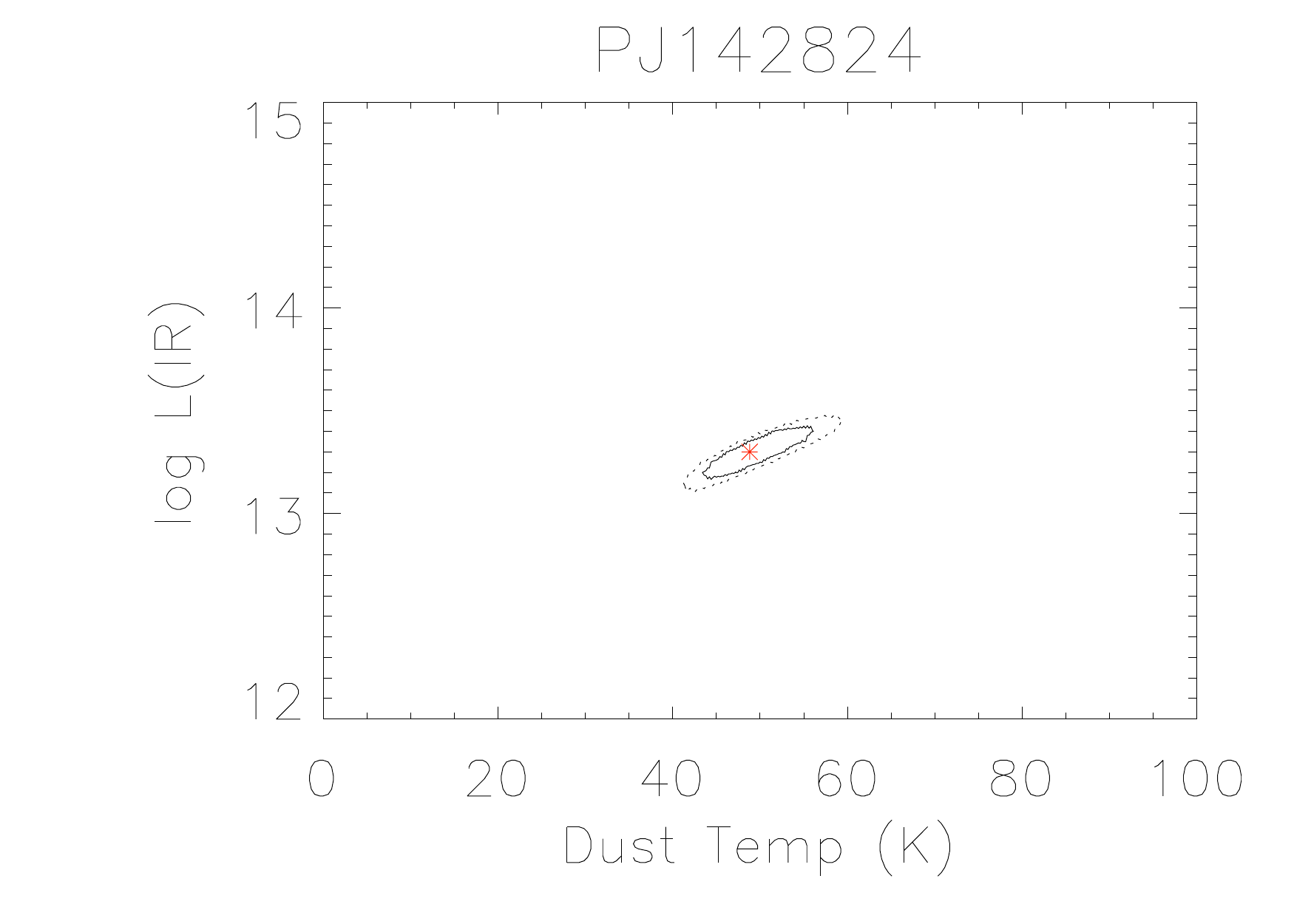}
\includegraphics[width=5.cm]{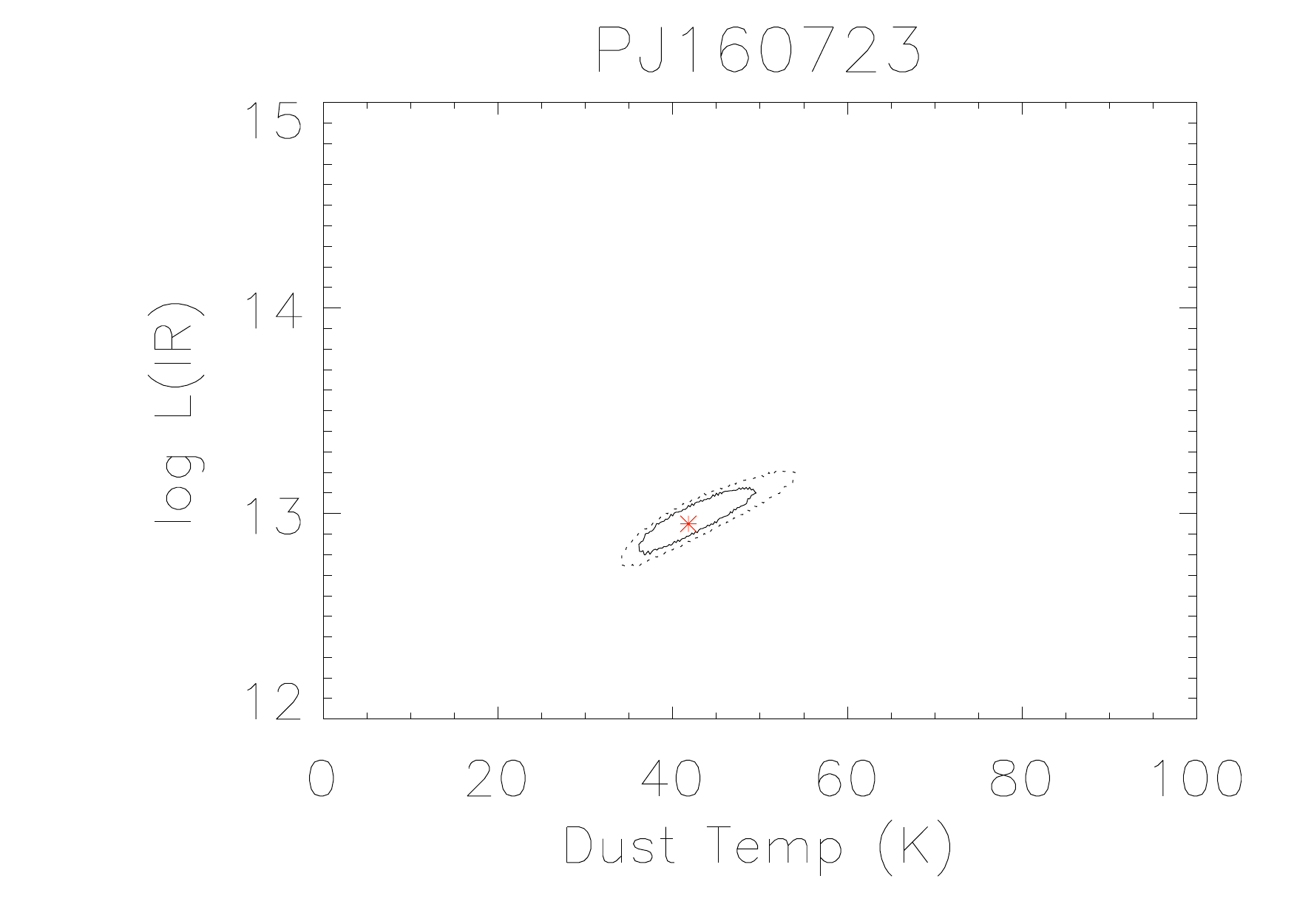}
\includegraphics[width=5.cm]{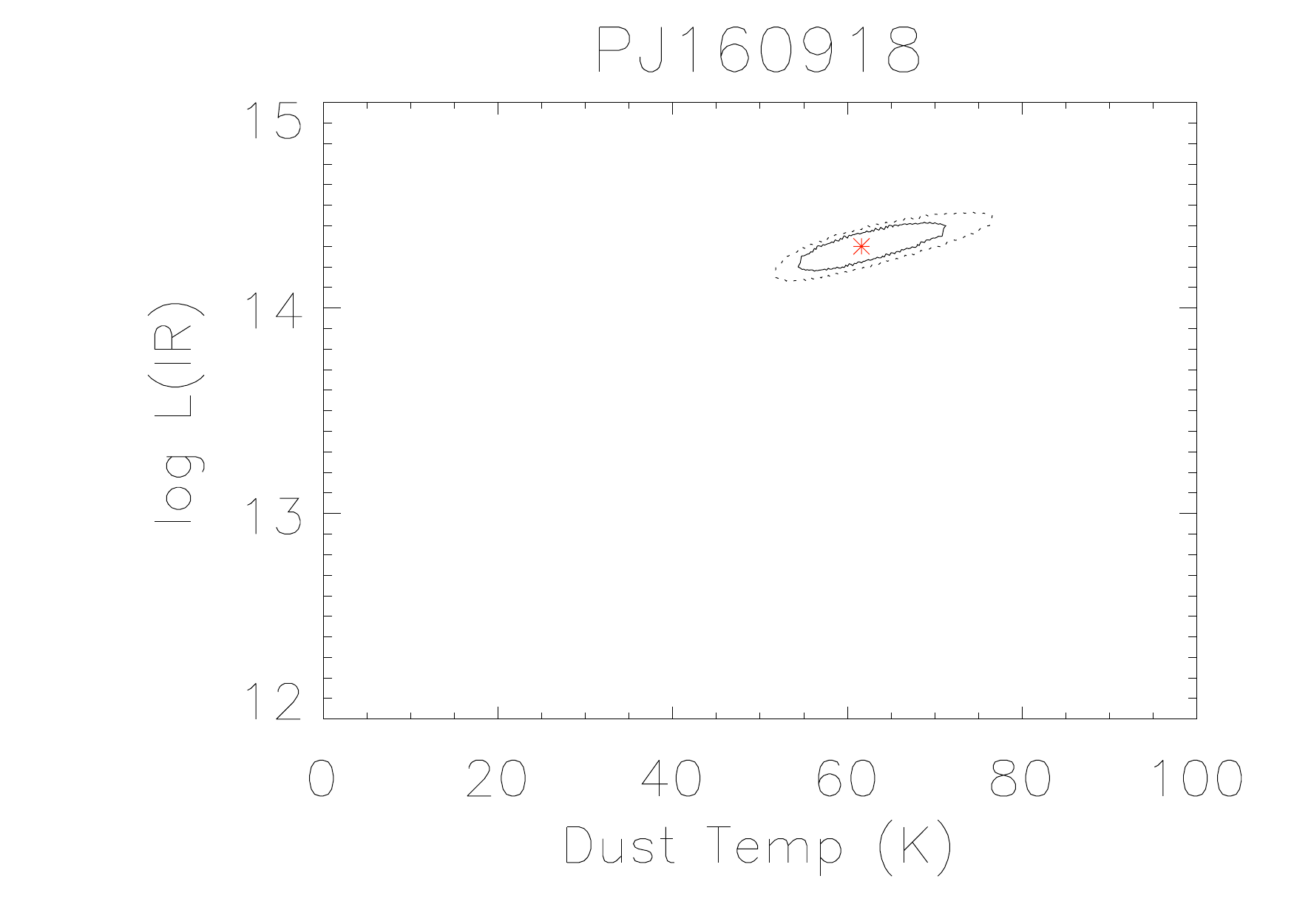}
\caption{Best fit $\chi^2$ analysis for the modified blackbody models.  The minimum $\chi^{2}$ is marked with an asterisk while the 68 and 90 percentile confidence levels are shown by a solid and dot-dash contours \citep{wall96}.  A weak constraint on $T_d$ is shown by the elongation of the confidence contours, and additional photometry measurements just shortward of the dust peak will be essential for an improved characterization of $T_d$.}
\label{fig:modifiedBB}
\end{figure*}

The resulting SED analyses for each sample galaxy are shown in Figure~\ref{fig:modifiedBB}.  The minimum $\chi^{2}$ is marked with an asterisk, and the 68 and 90 percentile confidence levels computed following  \citet{wall96} are shown by a solid and dot-dash contours.  By modifying the \citet{yun02} relation as adopted here, the term ``$SFR$" is no longer a true estimate of star formation rate and is simply a scaling factor instead.  We explicitly compute IR luminosity ($L_{IR},$ $\lambda =$ 8-1000 \micron) from the best fit model and compute star formation rate using the calibration by \citet{kennicutt98} corrected for the Kroupa IMF.  The best fit modified blackbody model parameters and derived quantities are reported in Table~\ref{tab:chisq}.

\section{Dust Mass Calculation}
\label{sec:Md}
	
	Following \citet{hildebrand83}, the total dust mass detected with a flux density of $S_\nu$ at the observed frequency $\nu$, located at a luminosity distance of $D_L$ is 
	$$M_{d} = \frac{S_{\nu} \, D_{L}^{2}}{\kappa_{\nu} B_{\nu}(T)}$$
	where $\kappa_\nu$ is the mass absorption coefficient and $B_\nu$ is the Planck function, $$B_{\nu}(T) = \frac{2h\nu^3}{c^2} \frac{1}{e^{(h \nu)/kT} - 1}.$$
	Mass absorption coefficient $\kappa_\nu$ is a major source of uncertainty, arising from our poor understanding of the grain properties such as density, size, shape, etc. \citep[see][]{hughes97}.  Several authors reviewed this subject and produced new estimates with varying assumptions \citep[e.g.][]{james02,dunne03b}, all falling within a factor of 2 of the Hughes et al. value $\kappa_{800\mu}=0.15$ m$^{2}$ kg$^{-1}$.  Thus for our dust mass calculation we adopt  
	$$\kappa_{\nu} = 0.15\,[\frac{\nu}{\nu_{800\mu}}]^{\beta}{\rm m}^2\,{\rm kg}^{-1}$$ where dust emissivity $\beta = 1.8$ is the same value used for the ISM mass calculation in \S~\ref{sec:ISMmass}. We a caution that a systematic uncertainly of up to a factor of  2 may result from this choice of $\kappa_\nu$.
		
	For a galaxy at high redshift $z$, we can re-write the equation for dust mass in terms of the measured flux density $S_{\nu,obs}$ at the observed frequency $\nu_{obs} \equiv \nu_{rest}/(1+z)$ as  
	$$M_{d} = \frac{S_{\nu,obs} \, D_{L}^{2}}{(1+z) \, \kappa_{\nu,obs} B_{\nu,obs}(T)}.$$
 
\end{document}